\documentclass[11pt,a4paper]{amsart}
\usepackage[top=4cm,bottom=4cm,left=3cm,right=3cm]{geometry}
\usepackage{amsmath,amssymb,amsfonts,amsthm}
\usepackage{graphicx}
\usepackage{textcomp}
\usepackage[dvipsnames]{xcolor}
\usepackage{url}

\usepackage[utf8]{inputenc}
\usepackage[T1]{fontenc}
\usepackage[nolist]{acronym}

\usepackage{amsmath, amssymb, amsfonts, amsthm}
\usepackage[all]{xy}
\usepackage{hyperref}
\usepackage{doi}
\usepackage{url}
\usepackage{mathtools}
\usepackage{calc}  
\usepackage{enumerate}
\usepackage{enumitem}
\usepackage{float}
\usepackage{bbm}
\usepackage{verbatim}
\usepackage{anyfontsize}

\usepackage{pgfplots}
\pgfplotsset{compat=1.18}
\usepgfplotslibrary{groupplots}

\pgfplotsset{
  johnson axis/.style={
    width=0.47\textwidth,
    height=0.36\textwidth,
    xlabel={$\delta=d_\lambda/(\lambda n)$},
    ylabel={$\rho=w/n$},
    xmin=0,
    ymin=0,
    grid=major,
    samples=250,
    legend style={
      font=\small,
      draw=none,
      fill=white,
    },
    legend pos=north west,
    tick label style={font=\small},
    label style={font=\small},
    title style={font=\small},
  },
}

\pgfmathdeclarefunction{johnson}{2}{%
  \pgfmathparse{#2/(1+sqrt(max(0,1-#2/#1)))}%
}

\newcommand{\plotGS}[4]{%
  \pgfmathsetmacro{\GSaverage}
    {((#1)-1+(#3)*((#1)^(#2)-(#1)))/((#1)^(#2))}%
  \pgfmathsetmacro{\GSendpoint}{\GSaverage/(#3)}%
  \addplot[
    #4,
    thick,
    domain=0:\GSendpoint
  ]
    {johnson(\GSaverage,(#3)*x)};
  \addplot[
    #4,
    only marks,
    mark=o,
    mark options={solid,fill=white},
    forget plot
  ]
    coordinates {(\GSendpoint,\GSaverage)};
}

\usepackage{graphicx}
\graphicspath{{images/}{../images/}}
\usepackage{subcaption}
\usepackage{tabularx}
\usepackage{array}
\usepackage{wrapfig}
\usepackage{longtable}
\usepackage{makecell}
\usepackage{multirow, multicol}
\usepackage{ltablex}\keepXColumns

\newcolumntype{Y}{>{\centering\arraybackslash}X}
\newcolumntype{Z}{>{\scriptsize}Y}
\usepackage{rotating}

\usepackage{tikz}
\usetikzlibrary{shapes,arrows,positioning,arrows.meta,chains,positioning,quotes,fit}
\usepackage{blkarray}
\usepackage{tkz-euclide}
\usetikzlibrary{patterns} 
\usepackage{mathrsfs}
\usetikzlibrary{fadings}
\usetikzlibrary{shadows.blur}
\usetikzlibrary{spy}
\usetikzlibrary{decorations.pathreplacing}

\usepackage{nicematrix}

\usepackage{pgfplots}
\usepgfplotslibrary{colorbrewer}
\pgfplotsset{compat=1.17}

\usepackage{subfiles}
\usepackage{blindtext}
\usepackage{chngcntr}

    \usepackage{todonotes}
\usepackage{diagbox}
\usepackage{cleveref}

\usepackage{centernot}

\usepackage{algorithm}
\usepackage{algpseudocode}

\usepackage[final]{pdfpages}

\usepackage{appendix}

\newtheorem{theorem}{Theorem}

\newtheorem{lemma}[theorem]{Lemma}
\newtheorem{proposition}[theorem]{Proposition}
\newtheorem{corollary}[theorem]{Corollary}
\newtheorem{example}[theorem]{Example}

\theoremstyle{definition}

\newtheorem{definition}[theorem]{Definition}

\theoremstyle{remark}

\usepackage{natbib}
\title{New Results on the Subfield Metric}

\author[A. Kosovac]{Ana Kosovac$^1$}
\author[J. Plechinger]{Jeanette Plechinger$^1$}
\author[N. Spirovski]{Nikola Spirovski$^2$}
\author[V. Weger]{Violetta Weger$^1$}

\address{$^1$Technical University of Munich\\
        Germany 
}

\address{$^2$Massachusetts Institute of Technology\\
        United States 
}
\email{ana.kosovac@tum.de}
\email{jeanette.plechinger@tum.de}
\email{nikola28@mit.edu}
\email{violetta.weger@tum.de}

\begin{document}

\begin{abstract}
In this paper, we further develop the  theory of the recently introduced $\lambda$-subfield metric. We investigate extremal codes, classical bounds and  asymptotic densities in this setting.

We first study codes attaining a Singleton-type bound, called Maximum $\lambda$-subfield Distance (M$\lambda$D) codes.  We establish structural restrictions on their parameters  and extend the theory to sublinear codes. We then investigate the asymptotic density of M$\lambda$D codes in several parameter regimes and compare their behavior with that of MDS and MRD codes.

We then turn to a Plotkin-like bound, where we construct a family of constant-weight codes attaining the bound and classify all non-degenerate linear constant-weight codes for $\lambda>1$.

Finally, we complement the collection of classical coding-theoretic bounds available for the $\lambda$-subfield metric by deriving Griesmer-, Elias- and Johnson-type bounds.  

\end{abstract}

\maketitle

\section{Introduction}

The $\lambda$-subfield metric was introduced by Grassl, Horlemann and Weger in \cite{Grassl_2023}, motivated by asymmetric quantum error correction.
For a vector over $\mathbb F_{q^m}$, the $\lambda$-subfield weight distinguishes between nonzero coordinates contained in the base field $\mathbb F_q$, having weight~$1$, and coordinates outside the base field, which we weight with~$\lambda\geq 1$.
Hence, for $\lambda=1$ one recovers the Hamming metric, while for larger values of $\lambda$, errors outside the base field are penalized more.

While the $\lambda$-subfield metric is defined coordinate-wise (like the Hamming metric), its dependence on the distinguished subfield $\mathbb F_q\subseteq\mathbb F_{q^m}$ suggests a connection to the rank metric.
Indeed, the rank metric also exploits the $\mathbb F_q$-structure of $\mathbb F_{q^m}$: rather than merely counting which coordinates are nonzero, it measures the $\mathbb F_q$-dimension of the $\mathbb{F}_q$-linear space spanned by all coordinates.
It is therefore natural to ask, which features of the $\lambda$-subfield metric are inherited from the Hamming metric, which from the rank metric, and which are genuinely specific to the new metric.

Our first answer to this question comes from codes that are optimal with respect to a Singleton-like bound.
In the Hamming metric and  in the rank metric, the corresponding optimal codes are the well-known Maximum Distance Separable (MDS) and Maximum Rank Distance (MRD) codes, respectively.
We call codes attaining the Singleton-like bound in the $\lambda$-subfield metric from \cite{Grassl_2023} Maximum $\lambda$-subfield Distance (M$\lambda$D) codes and show that they fit naturally between these two classical families. More precisely, for codes of length $n$ over $\mathbb{F}_{q^m}$, we establish the hierarchy
$\mathrm{MRD}\subseteq \mathrm{M}\lambda\mathrm{D}\subseteq \mathrm{MDS},$ for $m \geq n$.

This first result extends observations from \cite{Grassl_2023}
 and suggests that the $\lambda$-subfield metric lies in between these two metrics.
 We further derive structural restrictions on their parameters and investigate how the M$\lambda$D property changes with different values of $\lambda$.

This naturally raises the question of how large the class of M$\lambda$D codes actually is.
The density of codes attaining a Singleton-like bound can vary considerably with the underlying metric.
In the Hamming metric, MDS codes are dense as the field size grows and sparse as the length of the code grows. On the other hand, the asymptotic behavior of MRD codes is more subtle and has only recently been understood:
While $\mathbb F_{q^m}$-linear MRD codes are dense  for $q$ growing \cite{ale}, the larger class of $\mathbb F_q$-linear MRD codes are neither dense nor sparse  \cite{heide2,BR19,heide,anina}.

For other metrics the situation can be even more restrictive: in the Lee metric, non-trivial linear codes attaining the corresponding Singleton-like bounds essentially do not exist \cite{LeePaper}, while Maximum Homogeneous Distance codes over finite chain rings were recently shown to consist essentially only of lifted MDS codes \cite{andreas}.
Thus, maximum-distance codes can range from abundant families to almost nonexistent ones, depending on the underlying metric.

Following the general framework for densities of codes developed in \cite{gruica}, we determine the density of linear and sublinear M$\lambda$D codes in several asymptotic regimes, such as the length $n,$ the base field size $q$ or the extension degree for the field of linearity $\ell$ tending to infinity.

The inclusions $\mathrm{MRD}\subseteq  \mathrm{M}\lambda\mathrm{D}\subseteq \mathrm{MDS} $ 
also allow us to ask a more refined question:  how large is the family of M$\lambda$D codes within the set of MDS codes and how many M$\lambda$D codes are MRD. 

We next turn to optimal codes for a Plotkin-like bound given in \cite{Grassl_2023}.
These are constant-weight codes in the $\lambda$-subfield metric.
The classification of such codes has recently been completed for other non-Hamming metrics, including the Lee metric \cite{LeePaper} and the homogeneous metric over finite chain rings \cite{andreas}.
For the $\lambda$-subfield metric, we construct a  family of constant-weight codes and following the characterization of constant weight codes by Wood \cite{WoodStructure}, we show that, for $\lambda>1$, every non-degenerate linear constant-weight code is equivalent to a repetition of this construction.

Finally, we extend the collection of classical bounds known for the $\lambda$-subfield metric.
Building on the Singleton-, Plotkin-, sphere-packing- and Gilbert--Varshamov-like bounds established in \cite{Grassl_2023}, we derive Griesmer-, Elias- and Johnson-like bounds.

The Johnson-like bounds provide another interesting comparison with the Hamming metric and the rank metric.
In the Hamming metric, the classical Johnson bound gives a list-decoding radius beyond half the minimum distance, ensuring that the list of closest codewords remains polynomial in size.
In contrast, the same approach in the rank metric does not work as shown in \cite{wachter2013bounds}.
For the $\lambda$-subfield metric, we obtain a non-trivial list-decoding radius by exploiting its additive structure and the framework in \cite{guruswami2001extensions}.
Thus, in this respect, the $\lambda$-subfield metric behaves more similarly to the Hamming metric than to the rank metric.

Together, these results show that the $\lambda$-subfield metric is not simply a weighted variant of the Hamming metric: its  theory combines features from the Hamming and the rank metric, while exhibiting its own extremal and asymptotic behavior.

The paper is organized as follows. In Section~\ref{sec:prelim}, we introduce the necessary notation and recall the Hamming, rank and $\lambda$-subfield metrics.
Section~\ref{sec:mld} studies linear and sublinear M$\lambda$D codes and Section~\ref{sec:density} investigates their asymptotic density. 
Section~\ref{sec:constantweight} characterizes all constant-weight codes. 
In Section~\ref{sec:johnson}, we derive and compare several Johnson-type bounds, while  Section~\ref{sec:bounds} establishes additional classical bounds in the $\lambda$-subfield metric. Finally, we conclude the paper with open problems and future directions
in Section~\ref{sec:conclusion}.
Parts of the results presented in this paper originated in the bachelor's theses of the first two authors and within the MISTI research exchange program of the third author.

\section{Preliminaries}\label{sec:prelim}

We begin by fixing the notation and recalling the basic notions used throughout the paper.

Let $q$ be a prime power and let $\mathbb{F}_{q^m}$ be the finite field with $q^m$ elements, with multiplicative group $\mathbb{F}_{q^m}^{\times}$.

We denote by $[n]$ the integers from 1 to $n$ and for a set $S$, we denote by $|S|$ its cardinality. We denote the identity matrix of size $n \times n$ by $\text{Id}_n$ and for a matrix $A$, we denote by $A^\top$ its transpose. $\mathcal{O}$ denotes the big O notation, that is, $f(n)=\mathcal{O}(g(n))$
if there exist constants $C>0$ and $n_0$ such that
$|f(n)|\leq C|g(n)|$ for all $n\geq n_0$.

Throughout this paper, we fix  $\ell$ to be a divisor of $m$ and denote by $s=m/\ell$.
We consider the  field extension $\mathbb{F}_{q^\ell}\subseteq \mathbb{F}_{q^m}$ with $ [\mathbb{F}_{q^m}:\mathbb{F}_{q^\ell}]=s.
$

A linear code of length $n$ and dimension $k$ over $\mathbb{F}_{q^m}$ is a
$k$-dimensional $\mathbb{F}_{q^m}$-subspace
$\mathcal{C}\subseteq \mathbb{F}_{q^m}^n$. We refer to such a code as an
$[n,k]_{q^m}$ code. More generally, an $\mathbb{F}_{q^\ell}$-linear
(or sublinear) code is a $k$-dimensional
$\mathbb{F}_{q^\ell}$-subspace of $\mathbb{F}_{q^m}^n$ and is denoted by
$[n,k]_{q^\ell/q^m}$.

For a matrix $G\in\mathbb{F}_{q^m}^{k\times n}$, we denote by
$\langle G\rangle_{\mathbb{F}_{q^\ell}}$ its row span over
$\mathbb{F}_{q^\ell}$. We call $G$ a generator matrix of a $\mathbb{F}_{q^\ell}$-linear code
$\mathcal{C}$ if $\mathcal{C}=\langle G\rangle_{\mathbb{F}_{q^\ell}}$.

The dual code $\mathcal{C}^\perp$ consists of all vectors which are orthogonal to all codewords, i.e.,
$$\mathcal{C}^\perp = \{ x \in \mathbb{F}_{q^m}^n \mid \langle x,c \rangle=0 \forall c \in \mathcal{C}\},$$ where $\langle \cdot , \cdot \rangle $ denotes the standard inner product. 

A linear isometry  on $\mathbb{F}_{q^m}^n$ for a weight $\text{wt}$ is a linear map, which preserves the weight, i.e., $\text{wt}(x)= \text{wt}(\phi(x))$ for all $x \in \mathbb{F}_{q^m}^n.$ If a code $ \mathcal{C}'$ is the image of a code $\mathcal{C}$ under such an isometry, i.e., $\varphi(\mathcal{C})=\mathcal{C}'$, then we call the two codes equivalent with respect to the weight $\text{wt}.$

We call a code {non-degenerate} if no coordinate is identically zero on the whole code.

We next recall the Hamming metric and the rank metric, which will serve as the main points of comparison for the $\lambda$-subfield metric.

\begin{definition}
For $x\in\mathbb{F}_{q^m}^n$, the Hamming weight is given by its number of nonzero  entries, i.e.,
\[
\text{wt}_H(x)=|\{i\in[n]\mid x_i\neq 0\}|,
\]
and the rank weight is given by
\[
\text{wt}_R(x)
=
\dim_{\mathbb{F}_q}
(\langle x_1,\ldots,x_n\rangle_{\mathbb{F}_q}).
\]
The induced distances between $x,y \in \mathbb{F}_{q^m}^n$ are then given by 
$d_H(x,y)=\text{wt}_H(x-y)$ and
$d_R(x,y)=\text{wt}_R(x-y)$.
\end{definition}

 For a $[n,k]_{q^m}$ linear code $\mathcal{C}$, one defines its minimum Hamming distance, respectively minimum rank distance, as
 $$d_H(\mathcal{C})= \min \{\text{wt}_H(x) \mid x \in \mathcal{C}, x \neq 0 \}, $$ respectively as $$d_R(\mathcal{C})= \min \{\text{wt}_R(x) \mid x \in \mathcal{C}, x \neq 0 \}.$$
 Whenever the code $\mathcal{C}$ is clear from the context, we may simply write $d_H,$ respectively $d_R$.

 In the Hamming metric, the well-known Singleton bound gives an upper bound on $d_H$ given the other parameters $n,k.$ 
\begin{theorem}[Singleton bound]
Let $\mathcal{C}$ be an $[n,k]_{q^m}$ code. Then
\[
d_H(\mathcal{C})\leq n-k+1.
\] 
\end{theorem}
Also in the rank metric, a Singleton-like bound has been established.
\begin{theorem}[Rank-metric Singleton-like bound]
Let $\mathcal{C}$ be an $[n,k]_{q^m}$ code. Then
\[
mk \leq \max\{n,m\}(\min\{n,m\} -d_R(\mathcal{C})+1).
\] 
\end{theorem}
In particular, if  $n\leq m$, we note that the rank-metric Singleton-like bound simplifies to
\[
d_R(\mathcal{C})\leq n-k+1.
\]

The codes attaining the corresponding Singleton bounds are known as
Maximum Distance Separable (MDS) and Maximum Rank Distance (MRD)
codes, respectively.

We now recall the $\lambda$-subfield metric introduced in
\cite{Grassl_2023}.

\begin{definition}
Let $\lambda\geq 1$ be a real number. The $\lambda$-subfield weight on $\mathbb{F}_{q^m}$ is
defined as
\[
\text{wt}_\lambda(a)
=
\begin{cases}
0, & a=0,\\
1, & a\in \mathbb{F}_{q}\setminus\{0\},\\
\lambda, & a\in \mathbb{F}_{q^m}\setminus \mathbb{F}_{q}.
\end{cases}
\]
We extend the weight to vectors  $x \in \mathbb{F}_{q^m}^n$, additively, i.e.,
\[
\text{wt}_\lambda(x)
=
\sum_{i=1}^n \text{wt}_\lambda(x_i),
\]
and the $\lambda$-subfield distance between $x,y \in \mathbb{F}_{q^m}^n$ is then given  by
\[
d_\lambda(x,y)=\text{wt}_\lambda(x-y).
\]
\end{definition}

Similar to the Hamming and the rank metric, the minimum $\lambda$-subfield distance of a linear code $\mathcal{C}$ is  defined as
\[
d_\lambda(\mathcal{C})
=
\min\{\text{wt}_\lambda(c)\mid c\in\mathcal{C},\ c\neq 0\}.
\]
If the minimum $\lambda$-subfield distance of $\mathcal{C}$ is known, we refer to it as a $[n,k,d_\lambda]_{q^m}$ code.

For $\lambda>1$, the $\mathbb F_{q^m}$-linear isometries
of the $\lambda$-subfield metric are coordinate permutations
and coordinatewise multiplication by elements of
$\mathbb F_q^\times$. Thus their group is
\[
(\mathbb F_q^\times)^n\rtimes S_n.
\]
For $\lambda=1$, the metric is the Hamming metric, whose
$\mathbb F_{q^m}$-linear isometry group is given by
\[
(\mathbb F_{q^m}^\times)^n\rtimes S_n.
\]

Clearly, we have that 
\[
d_H(\mathcal{C})
\leq
d_\lambda(\mathcal{C})
\leq
\lambda d_H(\mathcal{C})
\]
for every code $\mathcal{C}$.

This observation leads to a Singleton-like bound for the  $\lambda$-subfield metric, as stated in \cite{Grassl_2023}.

\begin{theorem}[Singleton-like bound in $\lambda$-subfield metric]
\label{thm:singletonbound}
Let $\lambda \ge 1$ be a real number and $\mathcal{C} \subseteq \mathbb{F}_{q^m}^n$ be a code (not necessarily linear)
with minimum $\lambda$-subfield distance $d_\lambda$. Then
\begin{equation*}
    \left\lfloor \frac{d_\lambda - 1}{\lambda} \right\rfloor
\le n - \left\lceil \log_{q^m}(|\mathcal{C}|) \right\rceil.
\end{equation*}
Thus, for linear codes of dimension $k$, we get
\begin{equation*}
\left\lfloor \frac{d_\lambda - 1}{\lambda} \right\rfloor\le n - k. 
\end{equation*}
\end{theorem}

In the case of linear codes, the Singleton-like bound can be further tightened as in \cite{Grassl_2023}:
\begin{theorem}
\label{thm:linearsingleton}
Let $\lambda \ge 1$ be a real number and $\mathcal{C} \subseteq \mathbb{F}_{q^m}^n$ be a linear code of dimension $k$
with minimum $\lambda$-subfield distance $d_\lambda$. Then
\begin{equation*}
d_\lambda \le \lambda (n - k) + 1.
\end{equation*}
\end{theorem}

A code attaining the above bound with equality is called a
maximum $\lambda$-subfield distance code, or M$\lambda$D code. 

In \cite{Grassl_2023}, the authors also provide the size of $\lambda$-subfield metric balls, i.e.,
  \[
 B_{\lambda}\bigl(r,n,q^m) = \left\{ x \in \mathbb{F}_{q^m}^n \mid  \text{wt}_\lambda(x) \le r \right\}.
\]

\begin{lemma}
\label{lem:volume}
Let $\lambda \geq 1 $ be an integer and $0 \le r \le \lambda n$ be a positive integer. We have that
\begin{equation*}
   V_\lambda(r,n,q^m) \coloneqq \left| B_{\lambda}(r,n,q^m) \right|
= \sum_{j=0}^{r}
\sum_{i=0}^{\left\lfloor \frac{j}{\lambda} \right\rfloor}
(q^m - q)^i \binom{n}{i} (q - 1)^{j - \lambda i}
\binom{n - i}{j - \lambda i}.
\end{equation*}
\end{lemma}
To allow also real values of $\lambda$, we may adapt the formula as follows:
\begin{lemma}\label{lem:volumeR}
Let $\lambda \geq 1$ be a real number and $0\leq r\leq \lambda n$. Then
$$
V_\lambda(r,n,q^m) =\sum_{a=0}^{\min\{n,\lfloor r/\lambda\rfloor\}}
\binom{n}{a}(q^m-q)^a  \sum_{b=0}^{\min\{n-a,\lfloor r-\lambda a\rfloor\}}
\binom{n-a}{b}(q-1)^b.
$$
\end{lemma}

\begin{proof}
Let $x\in\mathbb F_{q^m}^n$ have exactly $a$ coordinates in
$\mathbb F_{q^m}\setminus\mathbb F_q$ and $b$ nonzero coordinates in $\mathbb F_q$, thus
$
\text{wt}_\lambda(x)=\lambda a+b.
$
For fixed $a$ and $b$, there are
$$
\binom{n}{a}(q^m-q)^a
\binom{n-a}{b}(q-1)^b
$$
such vectors. Summing over all $a,b$ satisfying $\lambda a+b\leq r$
gives the claim.
\end{proof}

In \cite{Grassl_2023} it was observed that the $\lambda$-subfield metric is translation invariant, that is,  the cardinality of a ball of fixed radius is independent of its center.

As we will also study codes attaining the Plotkin-type bound in the $\lambda$-subfield metric from \cite{Grassl_2023} and derive analogues of the Elias and Griesmer bounds, we briefly recall their classical statements in the Hamming metric.

\begin{theorem}[Plotkin bound]
Let $\mathcal{C}$ be a $[n,k]_{q^m}$ linear code with minimum Hamming distance $d_H$. Then
$$
d_H \leq \frac{n q^{mk}}{q^{mk}-1}\left(1-\frac{1}{q^m}\right).
$$
\end{theorem}

Wood showed in \cite{WoodStructure} that all constant Hamming weight codes, i.e., codes attaining this Plotkin bound, are $r$ fold repetitions of simplex codes (that is we simply repeat  each coordinate $r$ times).

\begin{theorem}[Griesmer bound]
Let $\mathcal{C}$ be an $[n,k]_{q^m}$ linear code with minimum Hamming distance $d_H$. Then
$$
n\geq \sum_{i=0}^{k-1} \left\lceil \frac{d_H}{q^{mi}}\right\rceil.
$$
\end{theorem}

And finally, having established a Johnson bound, one can also obtain an  Elias-Bassalygo bound:
\begin{theorem}[Elias-Bassalygo bound]
Let $\mathcal{C}\subseteq\mathbb{F}_{q^m}^n$ be a linear code with minimum Hamming distance $d_H$ and let $0\leq r\leq n(1-1/q^m)$ satisfy
$r^2-2nr+\frac{d_Hn}{1-1/q^m}>0.$ 
Then
$$
|\mathcal C| \leq \frac{q^{mn}}{V_H(r,n,q^m)}
\frac{d_H}{d_H-2r+\frac{r^2}{n(1-1/q^m)}},
$$
where
$$
V_H(r,n,q^m) = \sum_{i=0}^{r}\binom{n}{i}(q^m-1)^i
$$
denotes the volume of a Hamming ball of radius $r$.
\end{theorem}

Finally, we briefly recall the Johnson bound, which provides a guaranteed list-decoding radius.
In the unique decoding setting we need the balls around the codewords of radius $t$ to be disjoint in order to be able to assign 
 every received word to at most one closest codeword.  
 On the other hand, in the list decoding setting, we consider balls of larger radius than $t$ and ask: 
 how large can the decoding radius be while keeping the number of closest codewords  to a received vector polynomial?
 
\begin{definition} Let $\mathcal{C}$ be an $[n,k]_{q^m}$ linear code.
 We say that $\mathcal{C}$ is $(w,L)$-list decodable if for all $y\in \mathbb{F}_{q^m}^{n}$ we have \[\bigl|\{\, c\in \mathcal{C} \mid d_H(y,c)\le w \,\}\bigr|\le L.\]
\end{definition}  
 In the Hamming metric, the following bound gives a decoding radius up to which the list size remains polynomial:
\begin{definition}[Johnson radius] Let $d,n$ be positive integers with $d\le n$. The \emph{Johnson radius} is given by\[
 J(n,d)= n\left(1-\sqrt{1-\frac{d}{n}}\right)= n-\sqrt{n(n-d)}.\]
\end{definition}

\begin{corollary}\label{cor2.16}
  Let $\mathcal{C}$ be an $[n,k]_{q^m}$ linear code with minimum Hamming distance $d_H$ and let $w<J(n,d_H)$. Then $\mathcal{C}$ is $(w,L)\text{-list decodable}$
  with $$L\le\frac{d_H-w}{d_H-2w+\frac{w^2}{n}}.$$
\end{corollary}

There  also exists a $q$-ary version of this bound, which will come in handy for our comparisons:
\begin{theorem}\label{theorem2.17}
    Let $\mathcal{C}$ be a  $[n,k]_{q^m}$ (not necessarily linear) code with minimum Hamming distance $d_H$. Then $\mathcal{C}$ is $\bigl(w, q^m d_Hn\bigr)-$list decodable, if 
    $$w<J_{q^m} (n,d_H)=n \frac{q^m-1}{q^m} \left(1-\sqrt{1-\frac{q^m}{q^m-1}\frac{d_H}{n}}\right).$$
\end{theorem}

In contrast, no analogous Johnson-type result is known for the rank
metric. In fact, rank-metric codes may have exponential list size
as soon as we go beyond half the minimum distance \cite{wachter2013bounds}.

One of the goals of this paper is to investigate to what extent
Johnson-type arguments can be transferred to the
$\lambda$-subfield metric.

\section{Maximum $\lambda$-Subfield Distance Codes}
\label{sec:mld}

The Singleton bound is one of the most fundamental bounds in coding
theory. Codes attaining the Singleton bound with equality are of
particular interest, since they maximize the minimum distance among all
codes with the same length and dimension. In the Hamming metric and the rank
metric this leads to the well-studied classes of MDS and MRD codes,
respectively. In the $\lambda$-subfield metric, the corresponding
extremal objects are the Maximum $\lambda$-subfield Distance
(M$\lambda$D) codes, first introduced in \cite{Grassl_2023}.

In this section, we investigate structural properties of linear and
sublinear M$\lambda$D codes. We derive restrictions on their
parameters, compare M$\lambda$D codes for different values of
$\lambda$, and establish connections to MDS and MRD codes.

\subsection{Linear M$\lambda$D codes}
Recall that a $[n,k]_{q^m}$ linear code $\mathcal{C}$ is called a
M$\lambda$D code,
if it attains the bound in Theorem~\ref{thm:linearsingleton}, i.e., if $d_\lambda(\mathcal{C})= \lambda(n-k)+1.$

The following result compares the M$\lambda$D property for different
values of $\lambda$.

\begin{theorem}
\label{thm:l1l2}
Let $\mathcal{C}$ be an $[n,k]_{q^m}$  linear code and let
$\lambda_1,\lambda_2\geq 1$ be real numbers.
Then the following hold:
\begin{itemize}
    \item[(i)] if $\lambda_1\geq \lambda_2$ and $\mathcal{C}$ is
    M$\lambda_1$D, then $\mathcal{C}$ is M$\lambda_2$D,
    \item[(ii)] if $\lambda_2\geq\lambda_1\geq n$ and $\mathcal{C}$ is
    M$\lambda_1$D, then $\mathcal{C}$ is M$\lambda_2$D.
\end{itemize}
\end{theorem}

\begin{proof} Let us start with some notation. For a nonzero codeword $c$ of $\mathcal{C}$, we denote by 
 $a(c)$  the number of
nonzero coordinates of $c$ outside $\mathbb{F}_q$ and by $b(c)$  the
number of nonzero coordinates of $c$ contained in $\mathbb{F}_q$, that is:  $\text{wt}_{\lambda}(c)=\lambda a(c)+b(c)$.

For (i), we assume that $\mathcal{C}$ is M$\lambda_1$D but not
M$\lambda_2$D. As $\mathcal{C}$ is not M$\lambda_2$D,  there must exist some
$\tilde c\in\mathcal{C}\setminus\{0\}$ with
$\text{wt}_{\lambda_2}(\tilde c)<\lambda_2(n-k)+1$.
Following our notation, we write $a=a(\tilde c)$ and $b=b(\tilde c)$. If $b=0$, then
$\text{wt}_{\lambda_2}(\tilde c)=  \lambda_2 a<\lambda_2(n-k)+1$ and  hence $a\leq n-k$. Similarly, in the $\lambda_1$-subfield metric, we get that $\text{wt}_{\lambda_1}(\tilde c)=\lambda_1 a\leq
\lambda_1(n-k)<\lambda_1(n-k)+1$, contradicting the M$\lambda_1$D
property. If $b\geq 1$, we can upper bound the $\lambda_1$-weight of $\tilde c$ as 
\[
\begin{aligned}
\text{wt}_{\lambda_1}(\tilde c)
&=\lambda_1 a+b=\frac{\lambda_1}{\lambda_2}(\lambda_2 a+b)
   +b\left(1-\frac{\lambda_1}{\lambda_2}\right)\\
   & < \frac{\lambda_1}{\lambda_2} (\lambda_2(n-k)+1) + b\left(1-\frac{\lambda_1}{\lambda_2}\right)\\
&= \lambda_1(n-k)+\frac{\lambda_1}{\lambda_2}
   +b\left(1-\frac{\lambda_1}{\lambda_2}\right) \\ 
   & \leq \lambda_1(n-k)+ \frac{\lambda_1}{\lambda_2}+\left(1-\frac{\lambda_1}{\lambda_2}\right).
\end{aligned}
\]
As we assumed that $\lambda_1 \geq \lambda_2$, we get that $1- \lambda_1/\lambda_2 \leq 0$ and thus $\text{wt}_{\lambda_1}(\tilde c) \leq \leq \lambda_1(n-k)+1,$ which 
again contradicts that $\mathcal{C}$ is M$\lambda_1$D.

For (ii), we assume that $\mathcal{C}$ is M$\lambda_1$D, that is for all nonzero codewords $c \in \mathcal{C}$ we have that $\text{wt}_{\lambda_1}(c) \geq \lambda_1(n-k)+1$. We first show that every nonzero codeword
$c\in\mathcal{C}$ satisfies $a(c)\geq n-k$. Indeed, if
$a(c)\leq n-k-1$, then, since $b(c)\leq n$, we get that 
\[
\text{wt}_{\lambda_1}(c)
=
\lambda_1 a(c)+b(c)
\leq
\lambda_1(n-k-1)+n
\leq
\lambda_1(n-k),
\]
contradicting that $\mathcal{C}$ is M$\lambda_1$D.

If $a(c)=n-k$, then the
M$\lambda_1$D property forces $b(c)\geq 1$, as 
$$\lambda_1(n-k)+1 \leq \text{wt}_{\lambda_1}(c) = \lambda_1 a(c) +b(c) = \lambda_1 (n-k) +b(c).$$ Hence, for the $\lambda_2$-subfield weight, we get that 
$$\text{wt}_{\lambda_2}(c)=\lambda_2(n-k)+b(c)\geq
\lambda_2(n-k)+1.$$ Similarly, if $a(c)>n-k$, then $a(c)\geq n-k+1,$ and
therefore
\[
\text{wt}_{\lambda_2}(c)
=
\lambda_2 a(c)+b(c)
\geq \lambda_2 a(c) \geq 
\lambda_2(n-k+1)
\geq
\lambda_2(n-k)+1.
\]
Thus every nonzero codeword has $\lambda_2$-subfield weight at least
$\lambda_2(n-k)+1$, which implies that  $\mathcal{C}$ is M$\lambda_2$D.
\end{proof}

The converse of Theorem~\ref{thm:l1l2}(i) is, however, false in general.

\begin{example}
\label{ex:differentlambda}
Let $q=m=2$ and let
$\mathbb{F}_4=\mathbb{F}_2(\alpha)$ with $\alpha^2=\alpha+1$.
Consider the $[3,2]_{q^m}$ code generated by
\[
G=
\begin{pmatrix}
1 & 0 & \alpha\\
0 & 1 & \alpha+1
\end{pmatrix}.
\]
For $\lambda_1=2$,  the minimum $\lambda_1$-subfield distance is $d_{\lambda_1}=3$, and as $d_{\lambda_1}=3 = \lambda_1(n-k)+1=2 \cdot 1 +1$, the code is M$\lambda_1$D. However, it is not M$\lambda_2$D for
$\lambda_2=3$, since the codeword
$\tilde c=(1,1,1)$ has $\text{wt}_{\lambda_2}(\tilde c)=3<4
=\lambda_2(n-k)+1$.
\end{example}

The next result gives a stronger parameter restriction for a fixed
$\lambda>1$.

\begin{theorem}
\label{thm:parameterrestrictionmld2}
Let $\lambda>1$ be a real number. If $(q^m-1)/(q-1)\leq n-k$, then no linear
M$\lambda$D code exists.
\end{theorem}

\begin{proof}
Let $N=(q^m-1)/(q-1)$ and choose representatives
$a_1,\ldots,a_N$ for the quotient
$\mathbb{F}_{q^m}^\times/\mathbb{F}_q^\times$, where we set $a_1=1$.

Let $\mathcal{C}$ be an $[n,k]_{q^m}$ linear code. Up to permutations of the coordinates, we may assume that $\mathcal{C}$ has a  generator matrix of the form
$G=(\mathrm{Id}_k \  A)$, for some $A \in \mathbb{F}_{q^m}^{k \times (n-k)}$. Let us write the first row as
\[
c=(1,0,\ldots,0,b_1,\ldots,b_{n-k}).
\]

If one of the entries $b_j$ is zero, with $j \in \{1, \ldots, n-k\}$, then
$\text{wt}_\lambda(c)\leq 1+ \lambda(n-k-1)<\lambda(n-k)+1$, which implies that $\mathcal{C}$ is not M$\lambda$D.

We can thus assume that $b_j \neq 0$ for all $j \in \{1, \ldots, n-k\}$. Similarly, if some $b_j$ lies in
$\mathbb{F}_q^\times$, then
\[
\text{wt}_\lambda(c) \leq \lambda(n-k-1)+2 <
\lambda(n-k)+1,
\]
again contradicting the M$\lambda$D property.

It remains to assume that all $b_j$ lie outside
$\mathbb{F}_q^\times$. 
Note that we have $n-k \geq N$ entries $b_j$, while there are only
$N-1$ nontrivial cosets of
$\mathbb{F}_q^\times$ in $\mathbb{F}_{q^m}^\times$. Thus, the pigeonhole
principle implies that there exist two indices $j_1, j_2 \in \{1, \ldots, n-k\}$ with  $j_1\neq j_2$ such that
$b_{j_1}$ and $b_{j_2}$ lie in the same coset. Thus, there exists an element 
$a\in\mathbb{F}_{q^m}^\times$ such that both 
$a^{-1}b_{j_1},a^{-1}b_{j_2}\in\mathbb{F}_q^\times$.
We can now consider the codeword $a^{-1}c$, given by 

\[
a^{-1}c=(a^{-1},0,\ldots,0,a^{-1}b_1,\ldots,a^{-1}b_{n-k}).
\]
Thus, $a^{-1}c$ has  one coordinate (being $a^{-1}$) of weight at most
$\lambda$, two coordinates of weight $1$, namely $a^{-1}b_{j_1}$ and $a^{-1}b_{j_2}$, and at most $n-k-2$
remaining coordinates of weight at most $\lambda$. This implies that 
\[
\text{wt}_\lambda(a^{-1}c)
\leq
\lambda+2+\lambda(n-k-2)
=
\lambda(n-k)+2-\lambda
<
\lambda(n-k)+1,
\]
which contradicts the M$\lambda$D property.
\end{proof}

Unlike for MDS and MRD codes, the M$\lambda$D property is not preserved by
duality.
 
\begin{example}
\label{ex:dualnotmld}
Let $q=m=2$ and let
$\mathbb{F}_4=\mathbb{F}_2(\alpha)$ with $\alpha^2=\alpha+1$.
Consider the $[3,1]_{q^m}$ code generated by
\[
G=
\begin{pmatrix}
1 & \alpha & 1+\alpha
\end{pmatrix}.
\]
Every nonzero codeword of $\mathcal{C}$ has
$\lambda$-subfield weight $2\lambda+1=\lambda(n-k)+1=\lambda(3-1)+1$, so
$\mathcal{C}$ is M$\lambda$D for every $\lambda\geq1$.

However, the dual code $\mathcal{C}^{\perp}$ has dimension $k'=n-k=2$, and
$(1,1,1)\in\mathcal{C}^{\perp}$. Hence
$d_\lambda(\mathcal{C}^{\perp})\leq 3$. For $\lambda>2$, this gives
\[
d_\lambda(\mathcal{C}^{\perp})\leq 3<\lambda+1
= \lambda(n-k')+1= 
\lambda(3-2)+1,
\]
hence $\mathcal{C}^{\perp}$ is not M$\lambda$D.
\end{example}

 Further comparisons with MDS and MRD codes are
discussed in Section~\ref{sec:comparison}.

\subsection{Sublinear M$\lambda$D codes}

We now extend the notion of M$\lambda$D codes to codes that are
linear over a proper subfield.

Recall that we fix $m=\ell s$ and let
$\mathcal{C}\subseteq \mathbb{F}_{q^m}^n$ be an
$[n,k]_{q^\ell/q^m}$ code. 

In case of a sublinear code, we can generalize the Singleton-like bound of Theorem \ref{thm:singletonbound} to:

\begin{corollary}\label{cor:subsingleton}
Let $\lambda \geq 1$ be a real number.
Let $\mathcal{C}\subseteq\mathbb{F}_{q^m}^n$ be an $[n,k]_{q^\ell/q^m}$ linear code with minimum $\lambda$-subfield distance $d_\lambda$. Then
    \begin{equation*}
    \left\lfloor \frac{d_\lambda - 1}{\lambda} \right\rfloor
\le n - \left\lceil \frac{\ell k}{m}\right\rceil.
\end{equation*}
\end{corollary}
This follows directly from Theorem \ref{thm:singletonbound}, as  $\mathcal{C}$ is an $\mathbb{F}_{q^\ell}$-vector space of
dimension $k$ and we have $ |\mathcal{C}|=(q^\ell)^k=(q^m)^{k/s}. $

Codes attaining this bound are called sublinear M$\lambda$D codes.

Unlike the $\mathbb{F}_{q^m}$-linear case, where we could tighten the Singleton-like bound to Theorem \ref{thm:linearsingleton}, we cannot tighten the bound further, e.g. one might be tempted to assume that 
\begin{equation}\label{eq:sublinsingl}
d_\lambda \leq 
\lambda\left(
n-\left\lceil\frac{k}{s}\right\rceil
\right)+1.
\end{equation}
However, this quantity is not an upper bound on the minimum $\lambda$-subfield distance, as the following example shows:

\begin{example}
\label{ex:sublinearlargedistance}
Let $q=m=2$ and  let $\mathbb{F}_4=\mathbb{F}_2(\alpha)$ with $\alpha^2=\alpha+1$. Let us consider the $\mathbb{F}_2$-linear  code generated by
\[
G=
\begin{pmatrix}
\alpha & \alpha & \alpha
\end{pmatrix}.
\]
Then $\mathcal{C}$ is a $[3,1,1]_{4}$ code with
$d_\lambda(\mathcal{C})=3\lambda$. Since
\[
3\lambda > 2\lambda+1 =\lambda\left(
3-\left\lceil\frac{1}{2} \right\rceil \right)+1
\]
for every $\lambda>1$,    Equation \eqref{eq:sublinsingl} fails.
\end{example}

Nevertheless, Equation \eqref{eq:sublinsingl}  completely characterizes sublinear M$\lambda$D codes.

\begin{corollary}
\label{cor:sublinearmldcriterion}
Let $\lambda \geq 1$ be a real number.
An $[n,k]_{q^\ell/q^m}$  linear code $\mathcal{C}$ is sublinear M$\lambda$D if
and only if
\[
d_\lambda(\mathcal{C})
\ge
\lambda
\left(
n-\left\lceil\frac{k}{s}\right\rceil
\right)+1.
\]
\end{corollary}

By Corollary~\ref{cor:subsingleton},
\[
\left\lfloor\frac{d_\lambda-1}{\lambda}\right\rfloor
\leq n-\left\lceil\frac{k}{s}\right\rceil,
\]
and equality holds if
\[
d_\lambda\geq
\lambda\left(n-\left\lceil\frac{k}{s}\right\rceil\right)+1.
\]

Many structural results from the linear setting extend directly, for example:

\begin{corollary}\label{cor:l1l2sub}
Let $\lambda_1,\lambda_2\ge 1$ be real numbers.

\begin{enumerate}
\item
If $\lambda_1\ge \lambda_2$ and $\mathcal{C}$ is a sublinear
M$\lambda_1$D code, then $\mathcal{C}$ is a sublinear
M$\lambda_2$D code.

\item
If $\lambda_2\ge \lambda_1\ge n$ and $\mathcal{C}$ is a sublinear
M$\lambda_1$D code, then $\mathcal{C}$ is a sublinear
M$\lambda_2$D code.
\end{enumerate}
\end{corollary}

The proof follows verbatim from
Theorem~\ref{thm:l1l2} by replacing $n-k$ with
$n-\left\lceil \frac{k}{s}\right\rceil$.

Finally, duality behaves similarly to the linear case due to the counterexample~\ref{ex:dualnotmld}.

 \begin{corollary}\label{cor:alllambda}
Let $\mathcal C$ be a (sub-)linear code of length $n$ and let $\lambda=n+1$. If $\mathcal C$ is an
M$\lambda$D code, then $\mathcal C$ is an M$\lambda'$D code for every real number 
$\lambda'\geq 1$.
\end{corollary}

\begin{proof}
 
If $\lambda'\leq n+1$, the claim follows from
Corollary~\ref{cor:l1l2sub}(i) and if $\lambda'\geq n+1$, then
$\lambda'\geq n$, and the claim follows from
Corollary~\ref{cor:l1l2sub}(ii).
\end{proof}

 \subsection{Comparison with MDS and MRD Codes}
\label{sec:comparison}

The $\lambda$-subfield metric shares several features with both the
Hamming metric and the rank metric. In this section, we compare M$\lambda$D
codes with MDS and MRD codes. In particular, we show that for $m \geq n$ 
\[
\{\text{MRD codes}\}
\subset
\{\text{M$\lambda$D codes}\}
\subset
\{\text{MDS codes}\},
\]
 where both inclusions are
strict in general.

We first compare M$\lambda$D codes with MDS codes.

\begin{theorem}
\label{thm:mldismds}
Let $\lambda \geq 1$ be a real number.
Let $\mathcal{C}\subseteq\mathbb{F}_{q^m}^n$ be a linear M$\lambda$D
code. Then $\mathcal{C}$ is MDS. Moreover,
\[
d_H(\mathcal{C})
=
\frac{d_\lambda(\mathcal{C})-1}{\lambda}+1.
\]
\end{theorem}

\begin{proof}
Since $\mathcal{C}$ is M$\lambda$D, we have that $
d_\lambda(\mathcal{C})
=
\lambda(n-k)+1. $
Suppose there exists a nonzero codeword $c\in\mathcal{C}$ with
$\text{wt}_H(c)\leq n-k$. Then $
\text{wt}_\lambda(c)\leq \lambda(n-k), $
contradicting the M$\lambda$D property. Hence $
d_H(\mathcal{C})\geq n-k+1 $  and $\mathcal{C}$ is MDS. Substituting
$n-k=(d_\lambda(\mathcal{C})-1)/\lambda$ in $d_H(\mathcal{C})=n-k+1$
gives the claimed formula.
\end{proof}

The same relation also holds for sublinear codes.
\begin{corollary} 
Let $\lambda \geq 1$ be a real number.
Let $\mathcal{C} \subset \mathbb{F}_{q^m}^n$ be an $[n,k]_{q^\ell/q^m}$ linear M$\lambda$D code with parameters such that $s= m/\ell$ divides $k$. Then $\mathcal{C}$ is a sublinear MDS code.
\end{corollary}

Thus, every M$\lambda$D code is MDS. The converse is false in general,
since the M$\lambda$D property is not preserved under duality whereas
the MDS property is. Hence the dual code of Example~\ref{ex:dualnotmld} is an example of MDS code, which is not M$\lambda$D.

On the other hand, the relation between M$\lambda$D and MRD codes depends on the
extension degree $m$.

\begin{theorem}
\label{thm:mrdismld} Let $\lambda \geq 1$ be a real number.
Let $\mathcal{C}$ be a linear MRD code with $m\geq n$. Then
$\mathcal{C}$ is M$\lambda$D.
\end{theorem}

\begin{proof}
Since $\mathcal{C}$ is MRD and $m \geq n$, we have that $
d_R(\mathcal{C}) = n-k+1. $
Hence every nonzero codeword of $\mathcal{C}$ contains at least
$n-k+1$ $\mathbb{F}_q$-linearly independent entries.
Thus, at least $n-k$ entries lie outside
$\mathbb{F}_q$ and there is at least one additional nonzero entry.
Therefore
\[
\text{wt}_\lambda(c) \geq \lambda(n-k)+1
\]
for every nonzero codeword $c\in\mathcal{C}$, and hence  $\mathcal{C}$ is M$\lambda$D.
\end{proof}

As a consequence, Gabidulin codes  \cite{gabidulin, delsarte} provide an infinite family of
M$\lambda$D codes for $m \geq n$. 

The corresponding implication also holds in the sublinear setting.

\begin{corollary}
    \label{cor:mrdismldsublinear} Let $\lambda \geq 1$ be a real number.
Let $\mathcal{C}$ be a sublinear MRD code with $m\geq n$. Then $\mathcal{C}$ is a sublinear M$\lambda$D code.
\end{corollary}

However, the converse of Theorem~\ref{thm:mrdismld} is false.

\begin{example}
Let $\mathbb{F}_{16}=\mathbb{F}_2(\alpha)$ with $\alpha^4=\alpha+1$.
Consider the code $\mathcal{C}\subseteq\mathbb{F}_{2^4}^4$ generated by

\[
G=
\begin{pmatrix}
1 & 0 & \alpha & \alpha+1\\
0 & 1 & \alpha^2 & \alpha^2+1
\end{pmatrix}.
\]

Note that $
d_\lambda(\mathcal{C}) = 2\lambda+1 = \lambda(n-k)+1, $ and hence $\mathcal{C}$ is an M$\lambda$D code.

However, $ d_R(\mathcal{C})=2<n-k+1=3, $ and hence $\mathcal{C}$ is not MRD.

\end{example}

Note that in \cite[Theorem 6.2]{Grassl_2023}, a statement analogous to Theorem~\ref{thm:mrdismld} was formulated without the assumption $m\geq n$. The next examples will show that this restriction is, however, necessary.


  In fact, in the following we provide an example with $m<n$ which is MRD but not M$\lambda$D, as well as an example, which is M$\lambda$D but not MRD.

 \begin{example}\label{ex:mrdnotmld}
     Consider $\mathbb{F}_{8}=\mathbb{F}_2(\alpha)$ with $\alpha^3=\alpha+1$ and the $\left[6,2\right]_8$ code generated by
   $$
   G=\begin{pmatrix}  1 & 0 & 1 & \alpha & \alpha^2 & \alpha \\
   0 & 1 & \alpha & 1 & \alpha & \alpha^2
   \end{pmatrix}.
   $$
    This code has   minimum distances $d_R=3$ and $d_\lambda=2\lambda+2$. Thus, it is an MRD code, but not an M$\lambda$D code.  
\end{example}

\begin{example}\label{ex:mldnotmrd}
     Consider $\mathbb{F}_{343}=\mathbb{F}_7(\alpha)$ with $\alpha^3=\alpha^2+3$ and the $\left[6,2\right]_{343}$ code generated by
   \[
   G=\begin{pmatrix}  1 & 0 & \alpha & \alpha + 1 & \alpha + 2 & \alpha + 3 \\
   0 & 1 & \alpha^2 & \alpha^2+2\alpha+1 & \alpha^2+4\alpha+4 & \alpha^2+6\alpha+2
   \end{pmatrix}.
   \]
    This code has  minimum distances  $d_R=2$ and $d_\lambda=4\lambda+1$. Thus, it is an M$\lambda$D code, but not an MRD code.  
\end{example}

\section{Density of M$\lambda$D Codes}
\label{sec:density}

The existence of extremal codes is a classical topic in coding theory.
Beyond existence, one may ask how frequently such codes occur among all
codes with fixed parameters. This question has received considerable
attention for MDS and MRD codes. It was shown that MDS codes are dense for $q$ growing and sparse for $n$ growing. The case of MRD codes is a bit more subtle, as we have to differentiate whether the codes are $\mathbb{F}_q$-linear or $\mathbb{F}_{q^m}$-linear. In \cite{gruica2022commoncomplementslinearsubspaces}, it was shown that for $\mathbb{F}_{q^m}$-linear codes, MRD codes are dense for $q$ growing and sparse for $n$ growing. However, $\mathbb{F}_q$-linear MRD codes are neither dense nor sparse \cite{heide2,BR19,heide,anina}.

In this section, we investigate the density of M$\lambda$D codes.
Following the framework developed for MDS and MRD codes in \cite{gruica}, we study the
proportion of M$\lambda$D codes among all codes as either $n$ or $q$ grow. We consider both linear and sublinear (i.e., $\mathbb{F}_{q^\ell}$-linear) 
codes, where we also consider the density for $\ell$ growing. We then compare the resulting behavior with the known asymptotic
theory of MDS and MRD codes and compute the ratio of M$\lambda$D codes within the set of MDS codes and the ratio of MRD codes within the set of M$\lambda$D codes for $q \to \infty$.

Following the notation in \cite{gruica}, let us denote the {volume of the ball of radius $r$} in $\mathbb{F}_{q^m}^n$  by
    \[ V_D(r,n,q^m) = 
\left| B_{D}(r,n,q^m) \right|=\left| \left\{ x \in \mathbb{F}_{q^m}^n  \mid d_D(x,0) \leq r \right\} \right|,
\]
where $D$ serves as a placeholder to describe a translation-invariant distance function on $\mathbb{F}_{q^m}^n$.

\begin{definition}[{\cite[Definition 1.3]{gruica}}]
The \textit{density of $\mathbb{F}_{q^\ell}$-linear codes of minimum distance $d_D(\mathcal{C})$ at least $d$} is defined as
    \begin{equation*}
\delta_D(d,S,n,\ell,q^m) = 
\frac{
\left| \left\{ \mathcal{C} \subseteq \mathbb{F}_{q^m}^n  \mid 
|\mathcal{C}| = S,\,
d_D(\mathcal{C}) \ge d,\,
\mathcal{C} \text{ is } \mathbb{F}_{q^\ell}\text{-linear}
\right\} \right|
}{
\left| \left\{ \mathcal{C} \subseteq \mathbb{F}_{q^m}^n  \mid 
|\mathcal{C}| = S,\,
\mathcal{C} \text{ is } \mathbb{F}_{q^\ell}\text{-linear}
\right\} \right|
}.
\end{equation*}
The cardinality $S=q^{\ell k}$ depends on the $\mathbb{F}_{q^\ell}$-dimension $k$ of $\mathcal{C}$. 
\end{definition}

Our goal is to understand the limiting behavior of these proportions
as one of the parameters tends to infinity while the remaining
parameters are fixed.

The asymptotic behavior turns out to depend strongly on the parameter
regime under consideration. 

\subsection{The regime $n\to\infty$}

We begin with the asymptotic behavior as the code length tends to
infinity.

\begin{theorem}\label{thm:lineardensityfixedrate}
    Let  $q$ be a prime power, $m\geq 1$ be an integer and $\lambda>1$ be a real number. Fix a rate  \(R \in (0,1)\) and for $n \geq 1$ set \(k(n) = \lfloor Rn \rfloor\) and $d_\lambda(n)=\lambda(n-k(n))+1$. Then
\[
\lim_{n \to \infty} \delta_\lambda(d_\lambda(n), q^{m k(n)}, n, m,q^m) =0.
  \]
\end{theorem}

\begin{proof}
We have that
\[
n-k(n)=n-\lfloor Rn \rfloor\geq n-Rn=(1-R)n\;\xrightarrow[n\,\to\,\infty]{}\; \infty.
\]
Hence, we can fix $N\in\mathbb{N}$ large enough such that 
\[
n-k(n)\geq \frac{q^m-1}{q-1}
\]
for all $n\geq N$. By Theorem~\ref{thm:parameterrestrictionmld2}, there is no $[n,k(n)]_{q^m}$ linear M$\lambda$D code if $n\geq N$. Thus,
\[
\delta_\lambda(d_\lambda(n), q^{m k(n)}, n, m,q^m)  =0 
\]
for all $n\geq N$ and the statement follows immediately.
\end{proof}

Note  that this result can also be obtained, by observing that M$\lambda$D codes are also MDS codes, which are sparse as $n \to \infty.$
 
The situation of sublinear M$\lambda$D codes is presumably more difficult. We restrict ourselves to the case  where $s=m/\ell$ divides $k$.

\begin{theorem}\label{thm:sublinearntoinfty}
Let $\ell,s, d$ be positive integers, let
$\lambda \geq 1$ be a real number, and let $q$ be a prime power.
Then
$$
\lim_{n\to\infty} \delta_\lambda ( \lambda   d +1, 
q^{\ell s(n-  d)}, n, \ell, q^m )=0.
$$
\end{theorem}

\begin{proof}
    Every $\mathbb{F}_{q^\ell}$-linear code with $k=s(n- {d})$ and minimum $\lambda$-subfield distance at least $\lambda   d +1$ satisfies that
    $$  {d}=n-\frac{k}{s}=n-\left\lceil\frac{k}{s}\right\rceil$$ which implies that $$\lambda   d +1 =\lambda\left(n-\left\lceil\frac{k}{s}\right\rceil\right)+1. 
$$
    Consequently, it is a sublinear M$\lambda$D code and thus it is a sublinear MDS code with
    \[
    d_H=n-\frac{k}{s}+1= {d}+1.
    \]
    For the density, this means that
    \[
   0\leq \delta_\lambda (\lambda  d +1,  q^{\ell s(n- {d})}, n , \ell, q^{\ell s})\leq \delta_H (   d +1,  q^{\ell s(n- {d})}, n,  \ell, q^{\ell s}).
    \]
    By \cite[Theorem 4.7]{gruica}, the right-hand side tends to zero as $n\to\infty$ and the claim on the asymptotic result follows.
\end{proof}
 \subsection{Conditional Density of M$\lambda$D Codes for $q\to\infty$}
For $q\to\infty$, MRD and MDS codes behave differently, which makes this case particularly interesting. For example, additive, i.e., $\mathbb{F}_q$-linear, MDS codes have previously been shown to be dense \cite{gruica}, while additive MRD codes are sparse for $q\to\infty$ \cite{gruica2022commoncomplementslinearsubspaces}.  For the $\lambda$-subfield metric, we can show, under certain parameter restrictions,   that M$\lambda$D codes are dense.

\begin{theorem}\label{thm:qtoinf}
Let $n,\ell,s,k$ be positive integers with $k\leq ns$, let
$\lambda \geq 1$ be an integer, and set $m=\ell s$ and
$$
 d(\lambda)=\lambda\left(n-\left\lceil\frac{k}{s}\right\rceil\right)+1.
$$
If
$$
\ell(k-1)+(1-\ell s)\left\lceil\frac{k}{s}\right\rceil<0,
$$
then
$$
\lim_{q\to\infty} \delta_\lambda ( d(\lambda), q^{\ell k}, n, \ell, q^m) =1.
$$
\end{theorem}

\begin{proof}
Let us set 
$c=\left\lceil\frac{k}{s}\right\rceil$ and $t=n-c,$ as then $ d(\lambda)-1=\lambda t$. Recall that by Lemma~\ref{lem:volumeR}, we have that 
$$
V_\lambda(  d(\lambda)-1, n,q^m) =\sum_{a=0}^{t}
\binom{n}{a}(q^m-q)^a
\sum_{b=0}^{\min\{n-a,\lfloor\lambda(t-a)\rfloor\}}
\binom{n-a}{b}(q-1)^b.
$$
For every pair $(a,b)$ occurring in the sum, we have $b\leq n-a$.
Thus
$$
ma+b \leq ma+n-a =(m-1)a+n \leq (m-1)t+n.
$$
Since $t=n-c$, this gives $ma+b \leq m(n-c)+c.$

As all parameters except $q$ are fixed, it follows that
$$
V_\lambda(  d(\lambda)-1, n,q^m) = O\!\left( q^{m(n-c)+c} \right)
= O\!\left( q^{\ell ns-\ell s\left\lceil\frac{k}{s}\right\rceil
+\left\lceil\frac{k}{s}\right\rceil}
\right).
$$
Consequently,
$$
\frac{V_\lambda( d(\lambda)-1, n,q^m) }{ q^{\ell(ns+1-k)}}
=O\!\left(q^{\ell(k-1)
+(1-\ell s)\left\lceil\frac{k}{s}\right\rceil}
\right).
$$
Since we assumed that $
\ell(k-1)+(1-\ell s)\left\lceil\frac{k}{s}\right\rceil<0,
$,  the quotient tends to zero as
$q\to\infty$.
The claim now follows   from \cite[Theorem~3.5]{gruica}.

\end{proof}

The condition of Theorem~\ref{thm:qtoinf} simplifies considerably in the linear and additive, i.e., $\mathbb{F}_q$-linear, settings. 
\begin{corollary}\label{cor:lineardensityq} 
Let $\lambda \geq 1$ be an integer.
Linear M$\lambda$D codes with $m>k$ are dense as $q\to\infty$. 
\end{corollary} 
In fact, for linear codes we have $\ell=m$ and $s=1$. The condition of Theorem~\ref{thm:qtoinf} becomes \[ m(k-1)+(1-m)k<0, \] which is equivalent to $k<m$. 

\begin{corollary}\label{cor:additivedensityq} 
Let $\lambda \geq 1$ be an integer.
Additive M$\lambda$D codes with $m>k$ are dense as $q\to\infty$. 
\end{corollary} 

Indeed, for additive codes we have $\ell=1$ and $s=m$. Since $k<m$, we obtain $\left\lceil\frac{k}{s}\right\rceil=1. $ Substituting this into the condition of Theorem~\ref{thm:qtoinf} yields $ (k-1)+(1-m)<0, $ which is equivalent to $k<m$. 

The previous theorem gives a sufficient condition in terms of the code parameters. The next result provides a simple sufficient condition involving the  parameter $\lambda$.

\begin{theorem}\label{thm:qtoinf2}
Let $n,\ell,s,k$ be positive integers with $k\leq ns$, set $m=\ell s$,
and let $\lambda \geq1$ be an integer with $\lambda\leq m$.
Set

$$
 d(\lambda)
=
\lambda\left(n-\left\lceil\frac{k}{s}\right\rceil\right)+1.
$$

Then

$$
\lim_{q\to\infty}
\delta_\lambda
\bigl( d(\lambda),q^{\ell k},n,\ell,q^m\bigr)
=1.
$$

\end{theorem}

\begin{proof}
Set $ t=n-\left\lceil\frac{k}{s}\right\rceil,$
so that $ d(\lambda)-1=\lambda t$.
By Lemma~\ref{lem:volumeR}, we get that 
$$
V_\lambda(  d(\lambda)-1, n,q^m) = \sum_{a=0}^{t}
\binom{n}{a}(q^m-q)^a
\sum_{b=0}^{\min\{n-a,\lfloor\lambda(t-a)\rfloor\}}
\binom{n-a}{b}(q-1)^b.
$$
For every pair $(a,b)$ occurring in the sum, we have that
$b\leq \lambda(t-a).$
Hence
$$
ma+b
\leq ma+\lambda(t-a)
=\lambda t+(m-\lambda)a.
$$
Since $\lambda\leq m$ and $a\leq t$, it follows that
$
ma+b\leq mt.
$
As all parameters except $q$ are fixed, we obtain
$$
V_\lambda ( d(\lambda)-1,n,q^m) = O(q^{mt}).
$$

Using $m=\ell s$ and
$\lceil k/s\rceil\geq k/s$, we have
$$
mt=m\left(n-\left\lceil\frac{k}{s}\right\rceil\right)
\leq  m\left(n-\frac{k}{s}\right)
= \ell(ns-k).
$$
Therefore
$$
V_\lambda ( d(\lambda)-1,n,q^m) = O\!\left(q^{\ell(ns-k)}\right),
$$
and thus,
$$
\frac{ V_\lambda( d(\lambda)-1,n,q^m) }{ q^{\ell(ns+1-k)}
}=O(q^{-\ell}),
$$
which tends to zero as $q\to\infty$.
The claim now follows from \cite[Theorem~3.5]{gruica}.
\end{proof}

\subsection{The regime $\ell\to\infty$}

We next consider increasing the size of the  subfield over which the code is linear.

\begin{theorem}
Let $n,s$ be positive integers, let $\lambda \geq1$ be a real number, and let $q$ be a prime power.
Fix $1\leq k\leq ns$ and set

$$
 d(\lambda)
=
\lambda\left(n-\left\lceil\frac{k}{s}\right\rceil\right)+1.
$$

Consider the sequence  $ \bigl(\mathbb F_{q^{\ell s}}^n\bigr)_{\ell\geq1}.
$
Then M$\lambda$D codes are dense as $\ell\to\infty$, i.e.,

$$
\lim_{\ell\to\infty}
\delta_\lambda
\bigl(
 d(\lambda),
q^{\ell k},
n,
\ell,
q^{\ell s}
\bigr)
=1.
$$

\end{theorem}

\begin{proof}
Set

$$
t=n-\left\lceil\frac{k}{s}\right\rceil
\qquad\text{and}\qquad
r= d(\lambda)-1=\lambda t.
$$

Note that since $n,k,s$ and $\lambda$ are fixed, so are $t$ and $r$.

Recall that by Lemma~\ref{lem:volumeR}, we have that 
$$
V_\lambda(r,n,q^m) = \sum_{a=0}^{t}
\binom{n}{a}(q^m-q)^a
\sum_{b=0}^{\min\{n-a,\lfloor\lambda(t-a)\rfloor\}}
\binom{n-a}{b}(q-1)^b.
$$
Since $q,n,\lambda$ and $t$ are fixed, the binomial coefficients, the factors
$(q-1)^b$, and the number of summands are all independent of $\ell$.
Moreover, every   index satisfies $a\leq t$. Hence
$$
V_\lambda(r,n,q^m) = O\!\left(q^{mt}\right).
$$
Using $m=\ell s$ and
$$
t=n-\left\lceil\frac{k}{s}\right\rceil
\leq n-\frac{k}{s},
$$
we obtain
$$
V_\lambda(r,n,q^m) = O\!\left(
q^{\ell s\left(n-\left\lceil\frac{k}{s}\right\rceil\right)}
\right)
= O\!\left(q^{\ell(ns-k)}\right).
$$
Thus,
$$
\frac{V_\lambda(r,n,q^m) }{ q^{\ell(ns+1-k)} }
= O(q^{-\ell}),
$$
which tends to zero as $\ell\to\infty$.
The claim follows from \cite[Theorem~3.6]{gruica}.
\end{proof}


\subsection{Relative Densities}

The inclusions
\[
\{\mathrm{MRD}\}
\subset
\{\mathrm{M}\lambda\mathrm{D}\}
\subset
\{\mathrm{MDS}\}
\]
naturally raise the question of how large these subclasses are inside
the bigger code classes.

Let $
\delta_R, 
\delta_\lambda, $ and $
\delta_H $
denote the densities of MRD, M$\lambda$D and MDS codes,
respectively.

Whenever the corresponding densities are nonzero, we have
\[
\frac{|\mathrm{M}\lambda\mathrm{D}|}
     {|\mathrm{MDS}|}
=
\frac{\delta_\lambda}{\delta_H},
\qquad
\frac{|\mathrm{MRD}|}
     {|\mathrm{M}\lambda\mathrm{D}|}
=
\frac{\delta_R}{\delta_\lambda}.
\]

The first quotient can be evaluated in the asymptotic regimes where
both MDS and M$\lambda$D codes are dense.

\begin{corollary}
Let $  n,m,s$ be positive integers, let $\lambda \geq1$ be a real number,
and let $q$ be a prime power. Assume that the hypotheses of
Theorem~\ref{thm:qtoinf} are satisfied.

Then the proportion of MDS codes that are M$\lambda$D satisfies
\[
\lim_{q\to\infty}
\frac{
\left|\left\{
\mathcal{C}\subseteq\mathbb{F}_{q^m}^n
\mid 
\mathcal{C}\text{ is M}\lambda\mathrm{D}
\right\}\right|
}{
\left|\left\{
\mathcal{C}\subseteq\mathbb{F}_{q^m}^n
\mid 
\mathcal{C}\text{ is MDS}
\right\}\right|
}
=1.
\]

\end{corollary}

\begin{corollary}
Let $ n,s$ be positive integers, let $\lambda \geq1$ be a real number,
and let $q$ be a prime power. Assume that $m\geq n$ and that both
M$\lambda$D codes and MRD codes are dense in the considered asymptotic
regime.

Then the proportion of M$\lambda$D codes that are MRD satisfies
\[
\lim_{ {\ell\to\infty}}
\frac{
\left|\left\{
\mathcal{C}\subseteq\mathbb{F}_{q^m}^n
\mid 
\mathcal{C}\text{ is MRD}
\right\}\right|
}{
\left|\left\{
\mathcal{C}\subseteq\mathbb{F}_{q^m}^n
\mid 
\mathcal{C}\text{ is M}\lambda\mathrm{D}
\right\}\right|
}
=1.
\] 
\end{corollary}

However, when $n\to\infty$, both MDS and M$\lambda$D codes are
 sparse. Consequently, the quotient
$
\frac{|\mathrm{M}\lambda\mathrm{D}|}
     {|\mathrm{MDS}|}
=
\frac{\delta_\lambda}{\delta_H} $
cannot be determined from the density results alone and determining this
relative density remains an interesting open problem.

\section{Constant-Weight Codes in the $\lambda$-Subfield Metric}
\label{sec:constantweight}

Constant-weight linear codes form an extremal class in several metric settings. In the Hamming metric, the linear constant-weight codes have been completely characterized, to be only $r$ fold repetitions of simplex codes \cite{WoodStructure}. A simplex code is generated by a matrix, having as columns all the representatives of $(\mathbb{F}_{q^m}^k \setminus \{0\})/\mathbb{F}_{q^m}^\times$. That is: each non-zero vector of $\mathbb{F}_{q^m}^k$ appears once, up to scalar multiples. 

For constant $\lambda$-subfield weight codes, we will instead take all non-zero vectors of $\mathbb{F}_{q^m}^k$ up to multiples of $\mathbb{F}_q.$  Since for $\lambda=1$, the $\lambda$-subfield metric reduces to the Hamming metric, we will restrict ourselves in this section to  $\lambda>1$.


Our starting point is the Plotkin-type bound derived in
\cite{Grassl_2023}. For this, we define the average $\lambda$-subfield weight of
the alphabet $\mathbb{F}_{q^m}$ to be 
$$
{\omega}_\lambda = \frac{1}{q^m} \sum_{x\in\mathbb{F}_{q^m}} \text{wt}_\lambda(x)
= \frac{q-1+\lambda(q^m-q)}{q^m}.
$$
The Plotkin-type bound for linear codes can then be written as follows.

\begin{theorem}[Plotkin-type bound {\cite{Grassl_2023}}]
\label{thm:plotkin-linear}
Let $\mathcal{C}\subseteq\mathbb{F}_{q^m}^n$ be a linear code. Then
$$
d_\lambda(\mathcal{C}) \leq \frac{|\mathcal{C}|}{|\mathcal{C}|-1} n{\omega}_\lambda.
$$
\end{theorem}


Within the proof of such Plotkin-type bounds, one uses that 
$$\sum_{c \in \mathcal{C}\setminus\{0\}} \text{wt}(c) \geq d(\mathcal{C})( \mid \mathcal{C} \mid -1),$$
which is attained by codes, where each non-zero codeword has constant weight $d,$ thus called constant-weight codes.
\begin{definition}
A linear code $\mathcal{C}\subseteq\mathbb{F}_{q^m}^n$ is called a \emph{constant $\lambda$-subfield weight code} if there exists a  $w>0$ such that
$ \text{wt}_\lambda(c)=w $ for every $c\in\mathcal{C}\setminus\{0\}$.
\end{definition}

For such a code, the minimum distance is equal to its nonzero constant weight. We now give an explicit construction of constant-weight codes for arbitrary dimensions.

Let us define 
$$
\mathcal{P}_{q}(m,k) = \left(
\mathbb{F}_{q^m}^{k}\setminus\{0\} \right) /\mathbb{F}_{q}^{\times},
$$
where two vectors are equal, if they differ by multiplication with a scalar in $\mathbb{F}_q^\times$. Thus 
$$
M \coloneqq |\mathcal{P}_{q}(m,k)| = \frac{q^{mk}-1}{q-1} .
$$

We can choose $u_1, \ldots, u_M$ as representatives of each class in $\mathcal{P}_{q}(m,k)$  and define $G_{m,k,q}$ as a matrix, having $u_1, \ldots, u_M$ as  columns, i.e., 
$ G_{m,k,q} \in \mathbb{F}_{q^m}^{k\times M}$.
We denote the  code generated by $G_{m,k,q}$ by $\mathcal{S}_{q,m,k}$.

\begin{theorem}
\label{thm:constantweightconstruction}
For every prime power $q$, positive integers $m,k$ and real $\lambda\geq1$, the code $\mathcal{S}_{q,m,k}$ is a linear $[n,k,d_\lambda]_{q^m}$ constant $\lambda$-subfield weight code with length $n=\frac{q^{mk}-1}{q-1}$ and  minimum distance $$
d_\lambda=q^{m(k-1)} \left(1+\lambda\frac{q^m-q}{q-1}\right).
$$
\end{theorem}

\begin{proof}
Let $v\in\mathbb{F}_{q^m}^k\setminus\{0\}$ and consider the codeword $c=vG_{m,k,q}$.
Its coordinates $c_i$ are of the form $\langle v, u_i \rangle$.

We can consider the nonzero $\mathbb{F}_{q^m}$-linear map
\[
\varphi_v:
\mathbb{F}_{q^m}^k
\longrightarrow
\mathbb{F}_{q^m},
\qquad
 u\longmapsto \langle v, u \rangle.
\]
The map $\varphi_v$ is surjective and every fiber has cardinality $q^{m(k-1)}$.

Moreover, the multiplication of $u$ by an element $\mu$ of $\mathbb{F}_q^\times$ multiplies $\varphi_v(u)$ by the same element, that is $\varphi_v(\mu u) = \mu \varphi_v(u)$. Thus, for every class $[x]\in \mathbb{F}_{q^m}^\times/\mathbb{F}_q^\times$ there are exactly $q^{m(k-1)}$ classes $[u]\in\mathcal{P}_{q}(m,k)$ for which $[\langle v, u\rangle]=[x].$

Note that there is exactly one class in $\mathbb{F}_{q^m}^\times/\mathbb{F}_q^\times$, which is  contained in $\mathbb{F}_q^\times$, i.e., $\mathbb{F}_q^\times$ itself. Hence $c$ has exactly $q^{m(k-1)}$ nonzero coordinates in $\mathbb{F}_q$.

The remaining
$$
\frac{q^m-1}{q-1}-1 = \frac{q^m-q}{q-1}
$$
nonzero classes consist entirely of elements outside
$\mathbb{F}_q$. Thus $c$ has
$$
q^{m(k-1)}\frac{q^m-q}{q-1}
$$
coordinates in $\mathbb{F}_{q^m}\setminus\mathbb{F}_q$.

Finally, the remaining
$$
\frac{q^{mk}-1}{q-1} - q^{m(k-1)} \frac{q^m-1}{q-1} = \frac{q^{m(k-1)}-1}{q-1}
$$
coordinates are zero.

Therefore every nonzero codeword has $\lambda$-subfield weight
$$
q^{m(k-1)} + \lambda q^{m(k-1)} \frac{q^m-q}{q-1},
$$
independently of the choice of $v$.
\end{proof}
These constant $\lambda$-subfield weight codes are indeed optimal for the Plotkin-type bound of Theorem \ref{thm:plotkin-linear}.

\begin{corollary}
\label{cor:constantweightplotkin}
The code $\mathcal{S}_{q,m,k}$ attains the Plotkin bound in Theorem~\ref{thm:plotkin-linear} with equality.
\end{corollary}

\begin{proof}
Since $|\mathcal{S}_{q,m,k}|=q^{mk}$ and $M=(q^{mk}-1)/(q-1)$, the Plotkin bound gives
\begin{align*}
\frac{|\mathcal{S}_{q,m,k}|}{|\mathcal{S}_{q,m,k}|-1} M {\omega}_\lambda
&=\frac{q^{mk}}{q^{mk}-1}\frac{q^{mk}-1}{q-1}\frac{q-1+\lambda(q^m-q)}{q^m}\\
&=q^{m(k-1)}\left(1+\lambda\frac{q^m-q}{q-1}\right),
\end{align*}
which is precisely the minimum distance from Theorem~\ref{thm:constantweightconstruction}.
\end{proof}

We note that, for $\lambda>1$, this construction produces  the smallest length constant $\lambda$-subfield weight codes. We next show, that all constant $\lambda$-subfield weight codes are  (up to linear $\lambda$-subfield isometries) $r$ fold repetitions of $\mathcal{S}_{q,m,k}$, hence giving a complete characterization.

\begin{definition}
A weight on $\mathbb{F}_{q^m}$ is said to have the \emph{extension property} if every linear weight-preserving map between two linear codes extends to a linear isometry of the entire ambient space. That is: for $\mathcal{C}, \mathcal{C}' \subseteq \mathbb{F}_{q^m}^n$ two linear codes,  if there exists a linear map $\mu: \mathcal{C} \to \mathcal{C}'$, which preserves the   weight, then there exists a linear isometry $\overline{\mu}: \mathbb{F}_{q^m}^n \to \mathbb{F}_{q^m}^n$, such that  $\overline{\mu}_{\mid \mathcal{C}}= \mu.$ 
\end{definition}

The extension property is the main ingredient in the structure theorem for linear constant-weight codes of Wood \cite{WoodStructure}. We first verify that it holds for the $\lambda$-subfield weight.

\begin{proposition}
\label{prop:lambdaextension}
Let $\lambda>1$. The $\lambda$-subfield weight has the extension property.
\end{proposition}

\begin{proof}
Due to \cite[Proposition~12]{WoodEP} it is enough to show that the matrix associated to the weight has full rank.
This matrix is indexed by the equivalence classes of nonzero elements under multiplication by elements of $\mathbb F_q^\times$, and its entries record the weight between representatives of these classes.
For the $\lambda$-subfield weight, two representatives from the same class give weight $1$, while representatives from different classes give weight $\lambda$.
Hence, the associated matrix is given by
 $$
A=
\begin{pmatrix}
1 & \lambda & \cdots & \lambda\\
\lambda & 1 & \cdots & \lambda\\
\vdots & \vdots & \ddots & \vdots\\
\lambda & \lambda & \cdots & 1
\end{pmatrix} \in \mathbb{R}^{N \times N},
$$
where $
N=\frac{q^m-1}{q-1},
$
and we can easily compute its determinant to be 
$$
\det(A)
=
(1-\lambda)^{N-1}\bigl(1+(N-1)\lambda\bigr),
$$
which is nonzero for $\lambda>1$.
Hence $A$ has full rank, and the extension property follows.
\end{proof}

\begin{theorem}
\label{thm:constantweightclassification}
Let $\lambda>1$ be a real number and let $\mathcal{C}\subseteq\mathbb{F}_{q^m}^n$ be a non-degenerate linear code of dimension $k$. 
Then $\mathcal{C}$ has constant nonzero $\lambda$-subfield weight if and only if it is equivalent, with respect to the $\lambda$-subfield metric, to an $r$-fold repetition of $\mathcal{S}_{q,m,k}$ for some positive integer $r$.

In particular, $n = r \frac{q^{mk}-1}{q-1}$ and every nonzero codeword has weight
$$
r  q^{m(k-1)}\left(1+\lambda\frac{q^m-q}{q-1}\right).
$$
\end{theorem}

\begin{proof}
By Proposition~\ref{prop:lambdaextension}, the $\lambda$-subfield weight satisfies the extension property.
The structure theorem for linear constant-weight codes \cite[Theorem~5.4]{WoodStructure} therefore applies. 
Together with the smallest-length constant $\lambda$-subfield weight code constructed in Theorem~\ref{thm:constantweightconstruction}, it follows that every non-degenerate linear constant-weight code is equivalent (under the $\lambda$-subfield metric) to a $r$ fold repetition of $\mathcal{S}_{q,m,k}$.

The converse follows immediately from Theorem~\ref{thm:constantweightconstruction}, since $r$ fold repetitions preserve the constant-weight property.
\end{proof}

Thus, for $\lambda>1$, the preceding theorem gives a complete classification of non-degenerate linear constant-weight codes in the $\lambda$-subfield metric.

\begin{example}
\label{ex:constantweightF4}
Let $q=m=2$ and write $\mathbb{F}_4=\mathbb{F}_2(\alpha)$ with $\alpha^2=\alpha+1$. For $k=2$, we have $n=\frac{2^4-1}{2-1}=15.$
Since $\mathbb{F}_2^\times=\{1\}$, the columns of the canonical generator matrix are simply all nonzero vectors of $\mathbb{F}_4^2$. For example, we may take
$$
G=
\left(
\begin{array}{*{15}{c}}
1 & \alpha & \alpha+1 &
0 & 0 & 0 &
1 & 1 & 1 &
\alpha & \alpha & \alpha &
\alpha+1 & \alpha+1 & \alpha+1
\\
0 & 0 & 0 &
1 & \alpha & \alpha+1 &
1 & \alpha & \alpha+1 &
1 & \alpha & \alpha+1 &
1 & \alpha & \alpha+1
\end{array}\right).
$$

Every nonzero codeword contains precisely three zero coordinates, four nonzero coordinates in $\mathbb{F}_2$, and eight coordinates in $\mathbb{F}_4\setminus\mathbb{F}_2$. Hence
$ d_\lambda=4+8\lambda. $
In particular, for $\lambda=2$ the code is a $[15,2,20]_{4}$ constant $\lambda$-subfield weight code.

Moreover, the average weight is given by
$$
{\omega}_2=\frac{1+2(4-2)}{4}=\frac{5}{4},
$$
and the Plotkin-type bound from Theorem \ref{thm:plotkin-linear} gives
$$
d_\lambda\leq\frac{16}{15}\cdot15\cdot\frac{5}{4}=20.
$$
Thus the code attains the Plotkin bound with equality.
\end{example}

Finally, we want to note that our construction is a subset of the Hamming-metric constant weight codes. In fact, if we set $Q=q^m$ and $r=(Q-1)/(q-1)$, then the code
$\mathcal{S}_{q,m,k}$ can be viewed as a scaled $r$-fold
repetition of the $Q$-ary simplex code of dimension $k$, which we denote by $\mathcal{S}^H_{Q,k}.$

Indeed, let $S^H$ be a generator matrix of the
$Q$-ary simplex code whose columns represent the points of
$\mathrm{PG}(k-1,Q)$, and let $\alpha_1,\ldots,\alpha_r$
represent the cosets of $\mathbb{F}_q^\times$ in
$\mathbb{F}_Q^\times$. Then the columns of
\[
\bigl[
\alpha_1S^H \mid
\alpha_2S^H \mid \cdots \mid
\alpha_rS^H
\bigr]
\]
form a complete set of representatives of
$\mathcal{P}_q(m,k)$. Hence, with this choice of representatives,
\[
\mathcal{S}_{q,m,k} =
\bigl\{
(\alpha_1c,\ldots,\alpha_rc)
 \mid  c\in\mathcal{S}^H_{Q,k}
\bigr\}.
\]

In particular, $\mathcal{S}_{q,m,k}$ also has constant
Hamming weight
\[
rQ^{k-1} = q^{m(k-1)}\frac{q^m-1}{q-1},
\]
since every nonzero codeword of
$\mathcal{S}^H_{Q,k}$ has Hamming weight $Q^{k-1}$.

\section{Johnson-Type Bounds in the $\lambda$-Subfield Metric}
\label{sec:johnson}

The Johnson bound is one of the fundamental tools in list decoding.
Besides its importance as a decoding radius, it is remarkable because
it admits several different proofs, each emphasizing a different
geometric or combinatorial aspect of the Hamming metric.

We consider four different approaches: a direct consequence of the
Hamming Johnson bound, a Plotkin-type argument, an embedding into a
Hamming space,  and the general Guruswami--Sudan argument
\cite{guruswami2001extensions}.

 For each approach, we get a different bound, which covers different parameter ranges. We then compare the respective Johnson radii in Section \ref{subsec:comp}.

\subsection{A Direct Consequence of the Hamming Johnson Bound}

We begin with a first Johnson-type bound that follows immediately from the corresponding result in the Hamming metric. Recall that, for every $x\in\mathbb{F}_{q^m}^n$,
$$
\text{wt}_{\lambda}(x) \leq\lambda \text{wt}_H(x).
$$
Consequently, every code $\mathcal{C}\subseteq\mathbb{F}_{q^m}^n$ satisfies
$$
d_H(\mathcal{C}) \geq \frac{d_\lambda(\mathcal{C})}{\lambda}.
$$

Let us denote by  
$$ B_\lambda(y,r,n,q^m) = \{x \in \mathbb{F}_{q^m}^n \mid d_\lambda(x,y) \leq r\}$$
the $\lambda$-subfield ball around $y \in \mathbb{F}_{q^m}^n$ of radius $r$, and similarly 
$$ B_H(y,r,n,q^m) = \{x \in \mathbb{F}_{q^m}^n \mid d_H(x,y) \leq r\}.$$

\begin{corollary}
\label{cor:johnson-hamming}
Let $\lambda \geq 1$ and $w$ be  real numbers. 
Let $\mathcal{C}\subseteq\mathbb{F}_{q^m}^n$ be a code with minimum $\lambda$-subfield distance at least $d_\lambda$. If
$$
w < n\left( 1-\sqrt{1-\frac{d_\lambda}{\lambda n}}\right),
$$
then $\mathcal{C}$ is $(w,L)$-list decodable in the $\lambda$-subfield metric, where
$$
L \leq \frac{d_\lambda-\lambda w}{d_\lambda-2\lambda w+\frac{\lambda w^2}{n}}.
$$
\end{corollary}

\begin{proof}
Since $d_H(\mathcal{C})\geq d_\lambda/\lambda$, the Hamming Johnson bound of Corollary \ref{cor2.16} applied with minimum distance $d_\lambda/\lambda$ shows that $\mathcal{C}$ is Hamming-list decodable up to radius
$$
w<n\left(1-\sqrt{1-\frac{d_\lambda}{\lambda n}}\right),
$$
with list size
$$
L\leq \frac{\frac{d_\lambda}{\lambda}-w}{\frac{d_\lambda}{\lambda}-2w+\frac{w^2}{n}}
=\frac{d_\lambda-\lambda w}{d_\lambda-2\lambda w+\frac{\lambda w^2}{n}}.
$$

Since $\text{wt}_H(x)\leq\text{wt}_\lambda(x)$ for every $x\in \mathbb{F}_{q^m}^n$, we get that 
$$
B_\lambda(y,w,n,q^m)  \subseteq B_H(y,w,n,q^m) 
$$
for every $y\in\mathbb{F}_{q^m}^n$. Therefore the same list-size bound holds for the $\lambda$-subfield metric.
\end{proof}
Clearly, if we set $\lambda=1$, the $\lambda$-subfield metric coincides with the Hamming metric and also the two Johnson bounds, i.e., Corollary \ref{cor2.16} and Corollary \ref{cor:johnson-hamming}, coincide.

\subsection{Plotkin argument}

Another approach is based on  using a nonlinear Plotkin-like bound using the maximum weight of a code. Hence, we first require such bound for the $\lambda$-subfield distance.

\begin{proposition}\label{corol:plotkin}
Let $\lambda \geq 1$ be a real number. 
Let $\mathcal{C}\subseteq\mathbb{F}_{q^m}^n$ be a (not necessarily
linear) code with minimum $\lambda$-subfield distance
$d_\lambda(\mathcal{C})$, and let
\[
W_\lambda
=
\max\{
\text{wt}_\lambda(c)
\mid
c\in\mathcal{C}
\}.
\]
If  $
d_\lambda(\mathcal{C})
-
2W_\lambda
+
\frac{W_\lambda^2}{\lambda n}
>0, $
then
\[
|\mathcal{C}|
\leq
\frac{d_\lambda(\mathcal{C})}
     {d_\lambda(\mathcal{C})
      -2W_\lambda
      +\frac{W_\lambda^2}{\lambda n}}.
\]
\end{proposition}
Note that the condition $d_\lambda(\mathcal{C})-2W_\lambda+\frac{W_\lambda^2}{\lambda n}>0, $ is never satisfied for a linear code, as we have that  $W_\lambda \geq d_\lambda$. However, for non-linear codes, this can be satisfied. For example, the non-linear code $$ \mathcal{C} =\{(1,1,\alpha), (\alpha+1,\alpha,0)\} \subseteq \mathbb{F}_4^3,$$
where $\mathbb{F}_4=\mathbb{F}_2(\alpha)$ with $\alpha^2=\alpha+1$ and $\lambda=2$ is such that $W_\lambda=4$ while $d_\lambda=6.$

\begin{proof}
Let
\[
T=\sum_{c\in\mathcal{C}}\text{wt}_\lambda(c)
\]
denote the total $\lambda$-subfield weight of the code.

Since every pair of distinct codewords has distance at least
$d_\lambda(\mathcal{C})$,
\[
|\mathcal{C}|\bigl(|\mathcal{C}|-1\bigr)
d_\lambda(\mathcal{C})
\leq
\sum_{x,y\in\mathcal{C}}
d_\lambda(x,y). 
\]

Writing the right-hand side coordinatewise gives
\[
\sum_{x,y\in\mathcal{C}}
d_\lambda(x,y)
=
\sum_{i=1}^n S_i,
\]
where $
S_i
=
\sum_{a,b\in A_i}
\text{wt}_\lambda(a-b), $
and $A_i$ denotes the multiset of symbols appearing in the $i$-th
coordinate of all codewords in $\mathcal{C}$.

For any $a,b\in\mathbb{F}_{q^m}$, we have
\[\mathrm{wt}_\lambda(a-b) \leq \mathrm{wt}_\lambda(a)+\mathrm{wt}_\lambda(b)
-\frac{\mathrm{wt}_\lambda(a)\mathrm{wt}_\lambda(b)}{\lambda}.
\]
Indeed, equality holds if either symbol is zero. If both are
nonzero and belong to $\mathbb{F}_q$, the left-hand side is
at most $1$, whereas the right-hand side is
$2-1/\lambda\geq 1$. Otherwise, the right-hand side equals
$\lambda$, which is the maximum possible symbol weight.

Summing this inequality over the multiset $A_i$ of symbols
in coordinate $i$ gives
\[
S_i \leq 2|\mathcal{C}|\,T_i-\frac{T_i^2}{\lambda} \] where  \[
T_i=\sum_{a\in A_i}\mathrm{wt}_\lambda(a).
\]

Summing over all coordinates yields
\[
\sum_{x,y\in\mathcal{C}}
d_\lambda(x,y) = \sum_{i=1}^n S_i \leq  2|\mathcal{C}|T - \frac1\lambda \sum_{i=1}^nT_i^2.
\]

Applying the Cauchy--Schwarz inequality

gives
\[
\sum_{x,y\in\mathcal{C}}
d_\lambda(x,y)\leq2|\mathcal{C}|T-\frac{T^2}{\lambda n}.
\]

Since every codeword has weight at most $W_\lambda\leq\lambda n$,
we have $T\leq|\mathcal{C}|W_\lambda\leq|\mathcal{C}|\lambda n$.
The function
\[
f(t)=2|\mathcal{C}|t-\frac{t^2}{\lambda n}
\]
is increasing on $[0,|\mathcal{C}|\lambda n]$. Therefore,
\[
\sum_{x,y\in\mathcal{C}}d_\lambda(x,y) \leq f(T) \leq f\bigl(|\mathcal{C}|W_\lambda\bigr)
= |\mathcal{C}|^2 \left(2W_\lambda-\frac{W_\lambda^2}{\lambda n}\right).
\]

Combining the upper and lower bounds for the total distance $\sum_{x,y \in \mathcal{C}} d_\lambda(x,y)$ gives
\[
|\mathcal{C}|
\bigl(
|\mathcal{C}|-1
\bigr)
d_\lambda(\mathcal{C})
\leq
|\mathcal{C}|^2
\left(
2W_\lambda
-
\frac{W_\lambda^2}{\lambda n}
\right),
\]
which immediately rearranges to
\[
|\mathcal{C}|
\leq
\frac{d_\lambda(\mathcal{C})}
     {d_\lambda(\mathcal{C})
      -2W_\lambda
      +\frac{W_\lambda^2}{\lambda n}}.
\]
\end{proof}

 With this bound, we can now provide a second Johnson-like bound.
\begin{theorem}\label{theorem:plotkin}
Let $\lambda \geq 1$ be a real number. 
Let $\mathcal{C}\subseteq \mathbb{F}_{q^m}^n$ be a not necessarily
linear code with minimum $\lambda$-subfield distance $d_\lambda$.
Then $\mathcal{C}$ is $(w,\lambda n d_\lambda)$-list decodable for every
\[
w
\leq
\lambda n
\left(
1-\sqrt{
1-\frac{d_\lambda}{\lambda n}
+\frac{1}{\lambda^2n^2}
}
\right).
\]
\end{theorem}

\begin{proof} We need to show that for any $y \in \mathbb{F}_{q^m}^n$ we have that $|B_\lambda(y,w,n,q^m) \cap \mathcal{C}| \leq \lambda n d_\lambda$.
By translation invariance of the $\lambda$-subfield metric, it is
enough to consider balls centered at $0$. Let
$\mathcal{L}=\mathcal{C}\cap B_\lambda(0,w,n,q^m ) $ and set
$L=|\mathcal{L}|$. Then $\mathcal{L}$ can be seen as a (not necessarily linear) code with minimum distance at least
$d_\lambda$ and all its codewords have $\lambda$-subfield weight at
most $w$.

Applying Proposition~\ref{corol:plotkin} to $\mathcal{L}$ hence gives
\[
L
\leq
\frac{d_\lambda}
{d_\lambda-2w+\frac{w^2}{\lambda n}},
\]
or equivalently, that
\[
L
\leq
\frac{\lambda n d_\lambda}
{\lambda n d_\lambda-2\lambda nw+w^2}.
\]

Thus, to show that $L \leq \lambda n d_\lambda,$ it suffices to ensure that
 $
\lambda n d_\lambda-2\lambda nw+w^2\geq 1,
$
which is equivalent to
$(\lambda n-w)^2
\geq
\lambda^2n^2-\lambda n d_\lambda+1. $
Since $d_\lambda\leq \lambda n$, the right-hand side is nonnegative.
Thus, the condition is satisfied whenever
\[w 
\leq
\lambda n-\sqrt{\lambda^2n^2-\lambda n d_\lambda+1},
\]
which is precisely the stated bound on $w$.
\end{proof}

\subsection{Reductions to the Hamming metric}
In the $\lambda$-subfield metric, an error contained in the extension
field contributes a weight of $\lambda$, whereas an error contained in
the base field contributes only weight~$1$. This suggests representing
each extension-field error by $\lambda$ Hamming errors, thereby relating
the $\lambda$-subfield metric to the classical Hamming metric.
This leads us to the third Johnson-like bound.

First, we need to introduce some notation.

Let $\lambda\geq 1$ be an integer. Fix a set $R\subseteq\mathbb{F}_{q^m}$ of representatives of the additive cosets of $\mathbb{F}_q$ in $\mathbb{F}_{q^m}$, with $0\in R$.
Every $x\in\mathbb{F}_{q^m}$ has a unique decomposition $x=b+r$, where $b\in\mathbb{F}_q$ and $r\in R$. Define \begin{align*} 
\sigma:\mathbb{F}_{q^m} & \to R, \\ 
b+r & \mapsto r,
\end{align*} 
and
\begin{align*} 
\psi:\mathbb{F}_{q^m} & \to \mathbb{F}_{q^m}^{\lambda}, \\ 
x & \mapsto 
(\underbrace{\sigma(x),\ldots,\sigma(x)}_{\lambda-1},x).
\end{align*}
Extending $\psi$ coordinatewise gives the injective map
\begin{align*} 
\phi:\mathbb{F}_{q^m}^n &\to \mathbb{F}_{q^m}^{\lambda n},
\\ x_1,\ldots,x_n &\mapsto (\psi(x_1),\ldots,\psi(x_n)).
\end{align*}

Since $\sigma(x)=\sigma(y)$ if and only if
$x-y\in\mathbb{F}_q$, we have
\[
d_H(\psi(x),\psi(y))
=
\begin{cases}
0, & x=y,\\
1, & x-y\in\mathbb{F}_q^\times,\\
\lambda, & x-y\notin\mathbb{F}_q.
\end{cases}
\]
That is $\phi$ is an isometric embedding:
\[
d_H(\phi(x),\phi(y))=d_\lambda(x,y)
\] for all $x,y\in\mathbb{F}_{q^m}^n. $
\begin{example}
Let $\mathbb{F}_4=\mathbb{F}_2(\alpha)$ with $\alpha^2=\alpha+1$, and let $\lambda=2$.
The additive cosets of $\mathbb{F}_2$ are  $\{0,1\}$ and $\{\alpha,\alpha+1\}$.
Thus, choosing $R=\{0,\alpha\}$ gives $\sigma(0)=0$ and $\psi(0)=(0,0), \sigma(1)=0$ and $\psi(1)=(0,1)$, $\sigma(\alpha)= \alpha$ and $\psi(\alpha)=(\alpha,\alpha)$ and finally $\sigma(\alpha+1)=\alpha$ and $\psi(\alpha+1)=(\alpha,\alpha+1). $
For instance, if we consider the vector $x=(1,\alpha+1)$ we get  $
\phi(1,\alpha+1)=(0,1,\alpha,\alpha+1). $
\end{example}
\begin{theorem}
\label{theorem3.5}
Let $\lambda \geq 1$ be an integer. 
Let $\mathcal{C}$ be an $[n,k, d_\lambda]_{q^m}$ linear code.
If
\[
w<
\lambda n
\frac{q^m-1}{q^m}
\left(
1-
\sqrt{
1-
\frac{q^m}{q^m-1}
\frac{d_\lambda}{\lambda n}
}
\right),
\]
then $\mathcal{C}$ is $(w,L)$-list decodable with
\[
L \leq \lambda n q^m d_\lambda.
\]
\end{theorem}

 \begin{proof}
The image $\phi(\mathcal{C}) \subseteq \mathbb{F}_{q^m}^{n\lambda}$ is a (not necessarily linear) code with minimum Hamming distance $d_\lambda$. Moreover, for every
received word $r\in\mathbb{F}_{q^m}^n$, we have that 
\[
\bigl|\mathcal{C}\cap B_\lambda(r,w)\bigr| = \bigl|\phi(\mathcal{C})\cap B_H(\phi(r),w)\bigr|.
\]
Thus, applying Theorem~\ref{theorem2.17} to $\phi(\mathcal{C})$ gives the result.
\end{proof}

We again get that for $\lambda=1$ the bound coincides with the Hamming-metric bound in Theorem \ref{theorem2.17}.
%

\subsection{A Guruswami--Sudan Johnson Bound}
 The classical Hamming-metric Johnson bound can be obtained by embedding the alphabet $\mathbb{F}_{q^m} $ into a Euclidean space, in such a way that distances between symbols are represented by inner products. Guruswami and Sudan \cite{guruswami2001extensions} use this geometric viewpoint to derive a Johnson-type bound which can also be applied to other metrics. We adapt their argument to weight functions on finite abelian groups and then specialize it to the $\lambda$-subfield weight.

We first recall the following  bounds on sets of unit vectors, corresponding to \cite[Lemmas~3 and~4]{guruswami2001extensions}.

\begin{lemma}\label{lemma:VectorNumberBound}
Let $u_1,\dots,u_M\in\mathbb{R}^N$ be unit vectors.
\begin{enumerate}
\item If $\langle u_i, u_j\rangle <0$ for all $i\neq j$, then $M\leq 2N$.
\item If there exists $\varepsilon>0$ such that
$\langle u_i, u_j \rangle<-\varepsilon$ for all $i\neq j$, then
\(        M\leq 1+\frac{1}{\varepsilon}.
   \)
\end{enumerate}
\end{lemma}
 
In order to embed our metric space into a space with an inner product, we will require the following proposition, yielding a sufficient condition: 

\begin{proposition}\label{prop:JonhsonEmbeddingCondition}
Let $G$ be a finite abelian group of size $a$ and let $w:G\to\mathbb{R}$ be a symmetric weight function, i.e.,
$w(g)=w(-g)$ for all $g\in G.$
 Let us define the average weight as 
$$
\omega=\frac{1}{a}\sum_{g\in G}w(g).
$$
If
$$
\sum_{g\in G}w(g)\chi(g)\in\mathbb{R}_{\leq0}
$$

for every nontrivial irreducible character $\chi$ of $G$, then  there exists a matrix 
$A\in\mathbb R^{a \times a} $ such that
$$
A^\top A=\bigl(\omega-w(x-y)\bigr)_{x,y\in G}
$$
and
$$
A(1,\dots,1)^\top=0.
$$
\end{proposition}

\begin{proof}
Let us first consider $A^\top A$. For this, we define the matrix 
$$
W=\bigl(\omega-w(x-y)\bigr)_{x,y\in G} \in \mathbb{R}^{a \times a} 
$$
and note that since $w(g)=w(-g)$, we get that the matrix $W$ is symmetric. Moreover,
$$
W(1,\dots,1)^\top=0,
$$
since every row of $W$ sums to
$$
a\omega-\sum_{g\in G}w(g)=0.
$$
Hence, it is enough to show that $W$ is positive semidefinite: in this case we may choose a real matrix $A$ satisfying $A^TA=W$, and the   condition $W(1, \ldots, 1)^\top=0$ implies
$A(1,\dots,1)^\top=0$.
To determine the eigenvalues of $W$,  we identify vectors indexed by $G$ with functions $f:G\to\mathbb{C}$. 
 Then matrix multiplication gives
\[ (Wf)(x) =\sum_{y\in G}\bigl(\omega-w(x-y)\bigr)f(y).
\]
Thus $W$ acts by convolution with the function
$h(g)=\omega-w(g)$.

Hence the characters of $G$ diagonalize $W$. For the trivial character, i.e., $\chi_0(g)=1$ for all $g \in G$, the corresponding eigenvalue is
$$
\sum_{g\in G}\bigl(\omega-w(g)\bigr)=0.
$$
For every nontrivial character $\chi$, we may use the  orthogonality of characters to get that 
$$
\sum_{g\in G}\bigl(\omega-w(g)\bigr)\chi(g) = -\sum_{g\in G}w(g)\chi(g),
$$
which is nonnegative by assumption.
Thus all eigenvalues of $W$ are nonnegative, and the claim follows.
\end{proof}

For finite fields, the character condition in Proposition~\ref{prop:JonhsonEmbeddingCondition} can be replaced by a condition involving hyperplanes.

\begin{proposition}\label{prop:JohnsonFieldEmbeddingCondition}
Let $q=p^h$ be a prime power and let $w:\mathbb{F}_q\to\mathbb{R}$ be a weight function which is definite and invariant under nonzero multiplication, i.e., 
$w(0)=0$ and $w(x)=w(ax)$ for all $a\in\mathbb{F}_p^\times.$
We then define the average weight on $\mathbb{F}_q$ to be 
$$
\omega=\frac{1}{q}\sum_{x\in\mathbb{F}_q}w(x).
$$
If
$$
\frac{p}{q}\sum_{x\in H}w(x)\leq\omega
$$
for every $\mathbb{F}_p$-hyperplane $H\subseteq\mathbb{F}_q$, then there exists a matrix $A\in\mathbb{R}^{q\times q}$ such that
$$
A^\top A=\bigl(\omega-w(x-y)\bigr)_{x,y\in\mathbb{F}_q}
$$
and
$$
A(1,\ldots,1)^\top=0.
$$
 \end{proposition}

\begin{proof}
The additive characters of $\mathbb{F}_q$ are of the form

$$
\chi_a(x)=\zeta_p^{\text{Tr}_{\mathbb{F}_q/\mathbb{F}_p}(ax)},
$$
where $\zeta_p=e^{2\pi i/p}$.
For $a\neq0$, the kernel of $\chi_a$ is an $\mathbb{F}_p$-hyperplane.
Using the invariance $w(x)=w(mx)$ for $m\in\mathbb{F}_p^\times$, we obtain
$$
\sum_{x\in\mathbb{F}_q}w(x)\chi_a(x)
=
\frac{1}{p-1}
\left(
-\sum_{x\in\mathbb{F}_q^\times}w(x)
+
p\sum_{\substack{x\in\mathbb{F}_q^\times\\ \chi_a(x)=1}}w(x)
\right).
$$

Thus, the assumed hyperplane inequality implies that this quantity is in $\mathbb{R}_{\leq 0}$ and we can apply Proposition~\ref{prop:JonhsonEmbeddingCondition}.
\end{proof}

We can now apply the geometric argument of \cite[Theorem~1]{guruswami2001extensions}, which we quickly recall here for completeness.

\begin{theorem}[Johnson bound]\label{thm:JohnsonBound}
Let $G$ be a finite abelian group of size $a$ and $w$ be a symmetric weight function, which satisfies the assumptions of Proposition~\ref{prop:JonhsonEmbeddingCondition}, and assume additionally that $w(0)=0$. Let us extend $w$ additively to $G^n$, i.e., $w(x)=\sum_{i=1}^n w(x_i)$. Let $\omega$ denote the average weight, i.e., 
$$
\omega=\frac{1}{a}\sum_{g\in G}w(g).
$$
Let $\mathcal C\subseteq G^n$ and let $w_{\max}>0$ and $d$ be real numbers such that $w(c)\leq w_{\max}$ for all $c\in\mathcal C$
and $w(c-c')\geq d $ for all distinct $c,c'\in\mathcal C$ where $0<d<n\omega$.
If
$$
w_{\max}
<
n\omega
\left(
1-\sqrt{1-\frac{d}{n\omega}}
\right),
$$
then $|\mathcal C|\leq 2na. $
Moreover, if $w_{\max}<n\omega$ and $d-2w_{\max}+\frac{w_{\max}^2}{n\omega}>0$ then

$$
|\mathcal C|
\leq
\frac{d}
{d-2w_{\max}+\frac{w_{\max}^2}{n\omega}}.
$$
 
\end{theorem}

\begin{proof}

Let us write $ d=n\omega(1-\delta),$ and $w_{\max}=n\omega(1-\gamma), 
$ for some real numbers $\delta$ and $\gamma.$

For each $g\in G$, let $a_g$ denote the column of the matrix $A$ from Proposition~\ref{prop:JonhsonEmbeddingCondition} indexed by $g$. Thus,

$$
\langle a_g, a_{g'} \rangle =\omega-w(g-g').
$$

For $c\in\mathcal C$, define

$$
u_c=(a_{c_1}\mid\cdots\mid a_{c_n})\in\mathbb R^{na}.
$$

Hence, we get that

$$
\langle u_c, u_{c'} \rangle =n\omega-w(c-c').
$$

For $\alpha\geq0$ and distinct $c,c'\in\mathcal C$, we then obtain
\begin{align*}
\langle u_c-\alpha u_0, u_{c'}-\alpha u_0 \rangle &= n\omega-w(c-c')
-\alpha\bigl(2n\omega-w(c)-w(c')\bigr)  +\alpha^2n\omega\\
&\leq n\omega\bigl(\delta-2\alpha\gamma+\alpha^2\bigr).
\end{align*}

Similarly, we get that
$$
\|u_c-\alpha u_0\|^2 \leq n\omega\bigl(1-2\alpha\gamma+\alpha^2\bigr),
$$
where $\|u\|=\sqrt{\langle u, u\rangle}$ denotes the Euclidean norm.

Suppose that $0<w_{\max}<n\omega$, so that $0<\gamma<1$.
For $0\leq\alpha<1$, the vectors $u_c-\alpha u_0$ are nonzero,
since $\|u_c\|=\|u_0\|=\sqrt{n\omega}$.
We may therefore define the unit vectors
\[
v_c=\frac{u_c-\alpha u_0}{\|u_c-\alpha u_0\|}
\in\mathbb{R}^{na}.
\]
Let $c, c'$ be distinct codewords. If $\delta-2\alpha\gamma+\alpha^2<0$, we get that 
\[
\langle v_c,v_{c'}\rangle
\leq
\frac{\delta-2\alpha\gamma+\alpha^2}{1-2\alpha\gamma+\alpha^2}
\]
Here we use that the numerator is negative and that
\[
0<
\|u_c-\alpha u_0\|\,
\|u_{c'}-\alpha u_0\|
\leq n\omega(1-2\alpha\gamma+\alpha^2).
\]

For the first bound, the assumption on $w_{\max}$ is equivalent to
$\gamma>\sqrt{\delta}$. Thus, if we set $\alpha=\sqrt{\delta}$, we get that
\[
\delta-2\alpha\gamma+\alpha^2
=2\sqrt{\delta}(\sqrt{\delta}-\gamma)<0
\]
and the unit vectors $v_c$, indexed by $c\in\mathcal{C}$,
have strictly negative pairwise inner products.
We can hence apply Lemma~\ref{lemma:VectorNumberBound} in $\mathbb{R}^{na}$ to get that
\[
|\mathcal{C}|\leq 2na.
\]

For the second bound, we observe that
\[
d-2w_{\max}+\frac{w_{\max}^2}{n\omega}
=n\omega(\gamma^2-\delta).
\]
Thus the assumption
$d-2w_{\max}+\frac{w_{\max}^2}{n\omega}>0$
implies $\gamma^2>\delta$.

If we now set $\alpha=\gamma$, we get that
\[
\langle v_c,v_{c'}\rangle
\leq
\frac{\delta-\gamma^2}{1-\gamma^2}
=-\varepsilon,\]
where $
\varepsilon=\frac{\gamma^2-\delta}{1-\gamma^2}>0.$
For every $0<\varepsilon'<\varepsilon$, these inner products
are strictly less than $-\varepsilon'$.
We can now apply Lemma~\ref{lemma:VectorNumberBound}, which gives that 
\[
|\mathcal{C}|
\leq 1+\frac{1}{\varepsilon}
=\frac{1-\delta}{\gamma^2-\delta}
=\frac{d}
{d-2w_{\max}+\frac{w_{\max}^2}{n\omega}}.
\]
 
\end{proof}


We now specialize this bound to the $\lambda$-subfield weight  on $\mathbb{F}_{q^m}$, where $\lambda\geq1$.
Its average weight is given by
\[
\omega_\lambda^\star
=\frac{(q^m-q)\lambda+(q-1)}{q^m}.
\]
To apply Theorem~\ref{thm:JohnsonBound}, we first need to verify the
hyperplane condition in
Proposition~\ref{prop:JohnsonFieldEmbeddingCondition}.

Let $q=p^h$ be a prime power. The weight $w_\lambda$ is positive definite and
is invariant under multiplication by $\mathbb{F}_p^\times$.
Let $H\subseteq\mathbb{F}_{q^m}$ be an $\mathbb{F}_p$-hyperplane, and let  $s=|H\cap\mathbb{F}_q|$.
Since
\[
\dim_{\mathbb{F}_p}(H\cap\mathbb{F}_q)\geq h-1,
\]
we have that $s\geq q/p$. Thus,
\[
\sum_{x\in H}w_\lambda(x)  =(s-1)+\lambda\left(\frac{q^m}{p}-s\right)
\leq  \frac{q}{p}-1+\lambda\frac{q^m-q}{p}.
\]
It follows that
\[
\frac{p}{q^m}\sum_{x\in H}w_\lambda(x)
\leq
\frac{q-p+\lambda(q^m-q)}{q^m}
<
\omega_\lambda^\star.
\]
Thus the required embedding exists, and
Theorem~\ref{thm:JohnsonBound} gives the following
  Johnson-type bound:
\begin{corollary}[Johnson-type bound for the $\lambda$-subfield metric]
\label{cor:johnsonGS} Let $\lambda \geq 1$ be a real number.
Let $\mathcal C\subseteq\mathbb F_{q^m}^n$ be a code with minimum
$\lambda$-subfield distance $0<d_\lambda<n \omega_\lambda$, where
$$
\omega_\lambda = \frac{(q^m-q)\lambda+(q-1)}{q^m}.
$$

If $$w < n \omega_\lambda
\left(1-\sqrt{1-\frac{d_\lambda}{n\omega_\lambda}}
\right),$$
then the code $\mathcal C$ is $(w,L)$-list decodable in the
$\lambda$-subfield metric with $$L\leq \min\left\{2nq^m, \frac{d}
{d-2w_{\max}+\frac{w_{\max}^2}{n\omega}}\right\}.$$

If we additionally have that
$$
d_\lambda-2w+\frac{w^2}{n\omega_\lambda}>0,
$$
then we get  the sharper bound
$$
L\leq \frac{d_\lambda}{d_\lambda-2w+\frac{w^2}{n\omega_\lambda}}.
$$
\end{corollary}

\begin{proof}
Let us fix a vector $y\in\mathbb F_{q^m}^n$ and consider
$$
\mathcal L = \mathcal C\cap B_\lambda(y,w,n,q^m).
$$

By the translation invariance of the $\lambda$-subfield metric, we get that the translated set $\mathcal L-y$ 
has $\lambda$-subfield weight at most $w$ and minimum  distance at least
$d_\lambda$.
Applying Theorem~\ref{thm:JohnsonBound} with
$G=(\mathbb F_{q^m},+)$ and $\omega=\omega_\lambda$ gives the claimed
bounds on $|\mathcal L|$. 
\end{proof}

If we set $\lambda=1$, we again recover the Hamming-metric Johnson bound from Theorem \ref{theorem2.17}.


\subsection{Comparison of the Johnson-Type Bounds}
\label{subsec:comp}

We now compare the Johnson-type decoding radii of the different bounds.
Although the list sizes are relevant, our purpose here is to determine how far each argument guarantees list decodability. For fixed  $q,m,\lambda$, as all four approaches provide polynomial list-size  bounds,  we will only focus on  the radii rather than comparing the different list-size estimates.

To ease the notation, we will write $Q=q^m$ and define $ \delta=\frac{d_\lambda}{\lambda n},$ and $ 
\rho=\frac{w}{n}. $
Thus $\delta$ denotes the relative minimum distance, normalized by  the maximum possible weight $\lambda n$, while $\rho$ denotes the decoding radius per coordinate. Hence, $\rho$ may
exceed $1$. 

Recall that the average $\lambda$-subfield weight in $\mathbb{F}_{Q}$ is given by
\[
\omega_\lambda =\frac{q-1+\lambda(Q-q)}{Q}.
\]

We denote the normalized radii obtained from Corollary~\ref{cor:johnson-hamming},
Theorem~\ref{theorem:plotkin},  Theorem~\ref{theorem3.5},  and
Corollary~\ref{cor:johnsonGS}  by $\rho_{\mathrm{Ham}}$, $\rho_{\mathrm{Plotkin}}$,
$\rho_{\mathrm{emb}}$,   and
$\rho_{\mathrm{GS}}$, respectively.

The normalized radii are given by
\begin{align*}
\rho_{\mathrm{Ham}}(\delta) &=1-\sqrt{1-\delta},\\
\rho_{\mathrm{Plotkin}}(\delta)  &=\lambda\left(1-\sqrt{1-\delta+\frac{1}{\lambda^2n^2}}\right),\\
\rho_{\mathrm{emb}}(\delta)&=\lambda\frac{Q-1}{Q}\left(1-\sqrt{1-\frac{Q}{Q-1}\delta}\right),\\
\rho_{\mathrm{GS}}(\delta)&=\omega_\lambda\left(1-\sqrt{1-\frac{\lambda\delta}{\omega_\lambda}}
\right).
\end{align*}

We note that Corollary~\ref{cor:johnson-hamming} applies for  $0<\delta\leq1$, while Theorem~\ref{theorem:plotkin} gives a nonnegative decoding radius for
$1/(\lambda^2n^2)\leq\delta\leq1$.
Theorem~\ref{theorem3.5} additionally requires $\lambda$ to be an integer and $0<\delta<(Q-1)/Q$. Corollary~\ref{cor:johnsonGS} requires $0<\delta<\omega_\lambda/\lambda$.

For a fixed $x>0$, let us consider the function 
\[
F_b(x) =b\left(1-\sqrt{1-\frac{x}{b}}\right)
=\frac{x}{1+\sqrt{1-x/b}}.
\]
This expression is strictly decreasing in $b$ for $b\geq x$.
If we set 
\[
\beta_\lambda=\lambda\frac{Q-1}{Q},
\]
then
\[
\beta_\lambda-\omega_\lambda
=\frac{(q-1)(\lambda-1)}{Q}\geq0,
\]
and we have
\[
\omega_\lambda\leq\beta_\lambda<\lambda.
\]
By setting $x=\lambda\delta$, the monotonicity of $F_b(x)$ then gives
\[
\lambda\rho_{\mathrm{Ham}}(\delta)<\rho_{\mathrm{emb}}(\delta)
\leq\rho_{\mathrm{GS}}(\delta).
\]

Since $\frac{1}{\lambda^2n^2}>0$, we have that
\[
\rho_{\mathrm{Plotkin}}(\delta)=\lambda\left(
1-\sqrt{1-\delta+\frac{1}{\lambda^2n^2}}\right)<\lambda\left(1-\sqrt{1-\delta}\right)
=\lambda\rho_{\mathrm{Ham}}(\delta).
\]

In particular, since $\lambda\geq1$, we get that 
\[
\rho_{\mathrm{GS}}(\delta)>\rho_{\mathrm{Ham}}(\delta).
\]

For non-integer values of $\lambda$, the embedding argument does not apply,
but the comparisons
\[
\rho_{\mathrm{GS}}(\delta)>\lambda\rho_{\mathrm{Ham}}(\delta)\geq\rho_{\mathrm{Ham}}(\delta),
\] and \[
\rho_{\mathrm{GS}}(\delta)>\rho_{\mathrm{Plotkin}}(\delta)
\]
are still valid.

Thus $\rho_{\mathrm{GS}}(\delta)$ is the largest of these four radii.
However, if $\delta \geq \omega_\lambda /\lambda$,   Corollary \ref{cor:johnsonGS} does not apply, while another bound may still be applicable.

 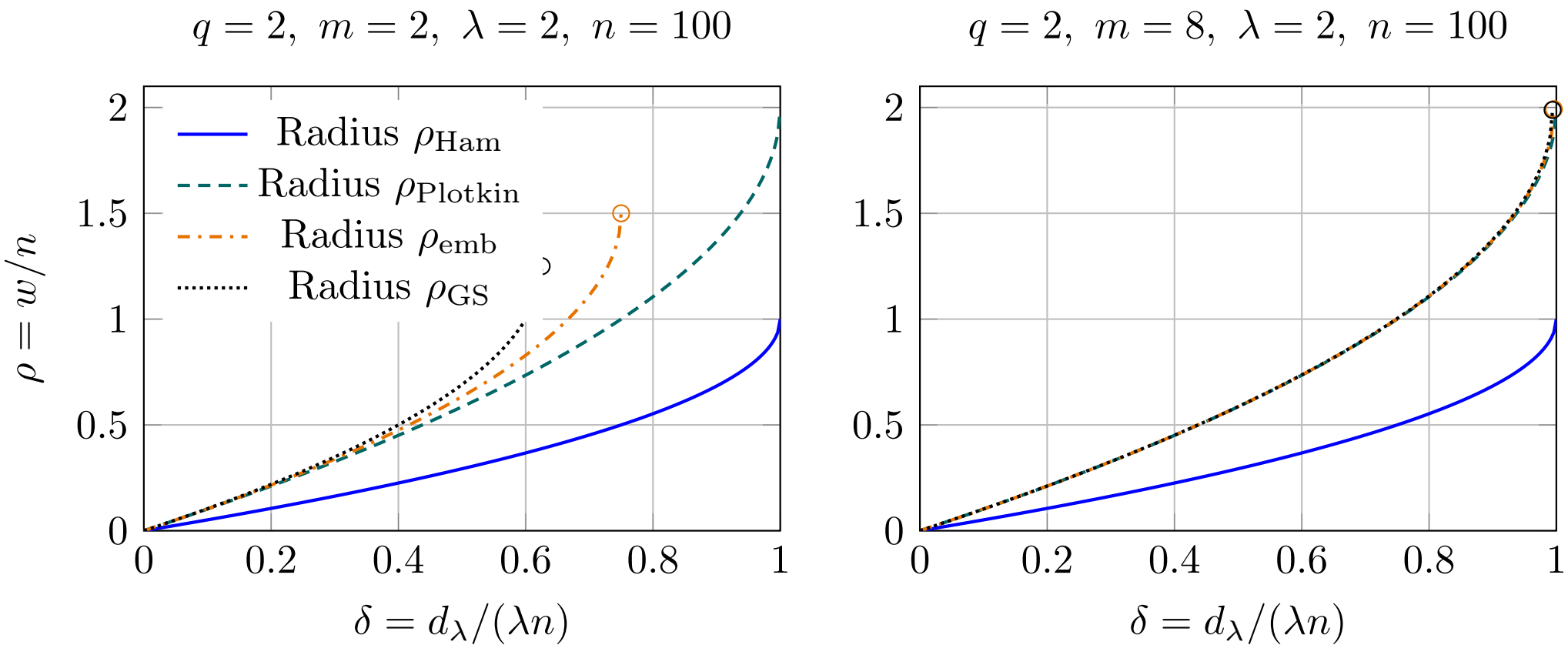
\begin{figure}[htbp]
\centering
\begin{tikzpicture}
\begin{groupplot}[
  johnson axis,
  group style={
    group size=2 by 1,
    horizontal sep=1.2cm,
  },
  xmax=1,
  ymax=2.1,
]

\nextgroupplot[
  title={$q=2,\ m=2,\ \lambda=2,\ n=100$},
]

\addplot[
  blue, thick,
  domain=0:1
]
  {johnson(1,x)};
\addlegendentry{Radius $\rho_{\text{Ham}}$}

\addplot[
  teal!80!black, thick, densely dashed,
  domain=0.000025:1
]
  {2*(1-sqrt(1-x+1/40000))};
\addlegendentry{Radius $\rho_{\text{Plotkin}}$}

\addplot[
  orange!90!black, thick, dashdotted,
  domain=0:0.75
]
  {johnson(1.5,2*x)};
\addlegendentry{Radius $\rho_{\text{emb}}$}

\addplot[
  orange!90!black,
  only marks, mark=o,
  mark options={fill=white},
  forget plot
]
  coordinates {(0.75,1.5)};

\plotGS{2}{2}{2}{black,densely dotted}
\addlegendentry{Radius $\rho_{\text{GS}}$}

\nextgroupplot[
  title={$q=2,\ m=8,\ \lambda=2,\ n=100$},
  ylabel={},
]

\addplot[
  blue, thick,
  domain=0:1
]
  {johnson(1,x)};

\addplot[
  teal!80!black, thick, densely dashed,
  domain=0.000025:1
]
  {2*(1-sqrt(1-x+1/40000))};

\addplot[
  orange!90!black, thick, dashdotted,
  domain=0:0.99609375
]
  {johnson(510/256,2*x)};

\addplot[
  orange!90!black,
  only marks, mark=o,
  mark options={fill=white},
  forget plot
]
  coordinates {(0.99609375,1.9921875)};

\plotGS{2}{8}{2}{black,densely dotted}

\end{groupplot}
\end{tikzpicture}
\caption{Comparison of the normalized Johnson radii.}
\label{fig:johnson-comparison-all}
\end{figure}

For the  parameters $q=2$, $m=8$ and $\lambda=2$,
 we have that  $\lambda=2,
\beta_\lambda=\frac{510}{256},$ and $
\omega_\lambda=\frac{509}{256}. $ 
As these values are close,   only two curves are distinguishable in the plot.

\section{Further Classical Bounds}\label{sec:bounds}

The original paper introducing the $\lambda$-subfield metric \cite{Grassl_2023} established several classical coding-theoretic bounds, including Singleton-, Plotkin-, sphere-packing and Gilbert--Varshamov-type bounds. In this section, we complement these results with a Griesmer- and an Elias-type bounds.

\subsection{A Griesmer-Type Bound}

We first relate the minimum $\lambda$-subfield distance of a linear code to its minimum Hamming distance.

\begin{lemma}
\label{lem:averageweight}
Let $w$ be a weight function on $\mathbb F_{q^m}$, which can be extended additively to $\mathbb{F}_{q^m}^n.$ Let us define its average weight on $\mathbb{F}_{q^m}$ as
$$
\omega = \frac{1}{q^m-1} \sum_{a\in\mathbb F_{q^m}^{\times}} w(a).
$$
Let $\mathcal{C} \subseteq \mathbb{F}_{q^m}^n$ be linear code  of minimum distance $d$, then
$$
d \leq \omega d_H(\mathcal C).
$$
\end{lemma}

\begin{proof}
Let $c\in\mathcal C\setminus\{0\}$ be a codeword of minimum Hamming weight $d_H(\mathcal C)$. 
By averaging over the $w$-weight for all $\alpha c$ over $\alpha\in\mathbb F_{q^m}^{\times}$ we get
$$
\frac{1}{q^m-1} \sum_{\alpha\in\mathbb F_{q^m}^{\times}} w(\alpha c)
= \omega d_H(\mathcal C).
$$
Hence at least one nonzero scalar multiple of $c$ has $w$-weight at most this average, i.e., for some $\alpha c \in \mathcal{C},$ we have that $w(\alpha c) \leq \omega d_H(\mathcal{C})$ and since $d \leq w(\alpha c)$ we get the claim. 
\end{proof}

For the $\lambda$-subfield weight, we observe  that averaging over all nonzero elements in $\mathbb{F}_{q^m}$ gives
$$
\omega_\lambda^\star = \frac{q-1+\lambda(q^m-q)}{q^m-1}.
$$
From this, we can get a Griesmer-type bound for the $\lambda$-subfield metric.
\begin{theorem}[Griesmer-type bound]
\label{thm:griesmerlambda}
Let $\mathcal C\subseteq\mathbb F_{q^m}^n$ be an $[n,k,d_\lambda]_{q^m}$ linear code. Then
$$
n \geq\sum_{i=0}^{k-1}\left\lceil\frac{d_\lambda}{q^{mi}\omega_\lambda^\star}\right\rceil .
$$
\end{theorem}

\begin{proof}
By Lemma~\ref{lem:averageweight}, we get that
$$
d_H(\mathcal C)\geq\frac{d_\lambda}{\omega_\lambda^\star}.
$$
Now we simply apply the classical Hamming-metric Griesmer bound over $\mathbb F_{q^m}$ to get 
$$n\geq\sum_{i=0}^{k-1}\left\lceil\frac{d_H(\mathcal C)}{q^{mi}}\right\rceil
\geq\sum_{i=0}^{k-1}\left\lceil\frac{d_\lambda}{q^{mi}\omega_\lambda^\star}\right\rceil .
$$
\end{proof}

\subsection{An Elias--Bassalygo-Type Bound}

We next show that the classical Elias--Bassalygo averaging argument in the Hamming metric extends directly to the $\lambda$-subfield metric. The essential ingredient is the translation invariance of the $\lambda$-subfield metric.
Recall that we denote by 
$$ B_\lambda(y,r,n,q^m) = \{x \in \mathbb{F}_{q^m}^n \mid d_\lambda(x,y) \leq r\}$$
the $\lambda$-subfield ball around $y \in \mathbb{F}_{q^m}^n$ of radius $r$ and that $V_\lambda(r,n,q^m)$ denotes the volume of a $\lambda$-subfield ball. 

\begin{lemma}[Elias--Bassalygo averaging]
\label{lem:eliasaveraging}
Let $\mathcal C\subseteq\mathbb F_{q^m}^n$ be a code and let $r\geq0$. 
Then there exists $y\in\mathbb F_{q^m}^n$ such that
$$
\left|\mathcal C\cap B_\lambda(y,r,n,q^m)\right|\geq \frac{|\mathcal C|\,V_\lambda(r,n,q^m)}{q^{mn}}.
$$
\end{lemma}

\begin{proof}
We consider the set
$$
\mathcal P=\left\{(c,y)\in\mathcal C\times\mathbb F_{q^m}^n \mid d_\lambda(c,y)\leq r \right\}
$$ and prove the statement using a double counting argument. 
When counting with respect to $c$, the translation invariance gives that
$$
|\mathcal P|=|\mathcal C|\,V_\lambda(r,n,q^m).
$$
On the other hand, when counting with respect to $y$, we get that 
$$
|\mathcal P|=\sum_{y\in\mathbb F_{q^m}^n}|\mathcal C\cap B_\lambda(y,r,n,q^m)|.
$$
Hence, the average intersection size is given by
$$
\frac{|\mathcal C|\,V_\lambda(r,n,q^m)}{q^{mn}},
$$
and therefore at least one center $y$ has  an intersection of at least
this size.
\end{proof}

Combining this averaging argument with an estimate for the $\lambda$-subfield volume gives an Elias-type upper bound.

\begin{theorem}[Elias-Bassalygo-type bound]
\label{thm:eliaslambda}
Let $\mathcal C\subseteq\mathbb F_{q^m}^n$ be a code of minimum $\lambda$-subfield distance $d_\lambda$. 
Suppose that every code of minimum distance at least $d_\lambda$ contained in a $\lambda$-subfield ball of radius $r$ has cardinality at most $L(r,d_\lambda)$. Then
$$
|\mathcal C|  \leq\frac{q^{mn}}{V_\lambda(r,n,q^m)} L(r,d_\lambda).
$$
\end{theorem}

\begin{proof}
By Lemma~\ref{lem:eliasaveraging}, there exists a center $y$ such that
$$
|\mathcal C\cap B_\lambda(y,r,n,q^m)|
\geq \frac{|\mathcal C|V_\lambda(r,n,q^m)}{q^{mn}}.
$$
The (not necessarily linear) subcode $\mathcal C\cap B_\lambda(y,r,n,q^m)$ still has minimum distance at least $d_\lambda$, and by assumption its cardinality is at most $L(r,d_\lambda)$. 

Thus
$$
\frac{|\mathcal C|V_\lambda(r,n,q^m)}{q^{mn}}\leq L(r,d_\lambda),
$$
which gives the claim.
\end{proof}

In particular, inserting any of the Johnson-type list-size bounds from Section~\ref{sec:johnson} into Theorem~\ref{thm:eliaslambda} yields an explicit Elias--Bassalygo-type bound for the $\lambda$-subfield metric. 
In particular, if we insert the list size of Corollary \ref{cor:johnson-hamming}, we get the following bound: 

Applying Corollary~\ref{cor:johnson-hamming} in
Theorem~\ref{thm:eliaslambda} gives the following explicit bound.

\begin{corollary}
\label{cor:eliaslambda-hamming}
Let $\lambda\geq1$ be a real number and let
$\mathcal C\subseteq\mathbb F_{q^m}^n$ be a code with
minimum $\lambda$-subfield distance $d_\lambda>0$.
Let $r$ be a real number with
\[
0\leq r< n\left(1-\sqrt{1-\frac{d_\lambda}{\lambda n}}\right).
\]
Then, 
\[
|\mathcal C| \leq
\frac{q^{mn}}{V_\lambda(r,n,q^m)} \frac{d_\lambda-\lambda r}{d_\lambda-2\lambda r+\frac{\lambda r^2}{n}}.
\]
\end{corollary}

\begin{proof}
By Corollary~\ref{cor:johnson-hamming}, every $\lambda$-subfield ball of radius $r$ contains at most
\[
\frac{d_\lambda-\lambda r}
{d_\lambda-2\lambda r+\frac{\lambda r^2}{n}}
\]
codewords of $\mathcal C$.
The result now follows from Theorem~\ref{thm:eliaslambda}.
\end{proof}

 \section{Conclusion}\label{sec:conclusion}

In this paper, we continued the study of the recently introduced $\lambda$-subfield metric by developing several aspects of its  theory. Our results cover extremal codes, classical bounds and asymptotic densities.

We first investigated Maximum $\lambda$-subfield Distance (M$\lambda$D) codes. Besides establishing structural restrictions and their relation to MDS and MRD codes, we introduced sublinear M$\lambda$D codes and studied their asymptotic density. 
We further constructed a family of constant-weight codes attaining the Plotkin bound and, using the extension theorem of Wood, obtained a classification of all non-degenerate linear constant-weight codes for $\lambda>1$.

We also complemented the collection of classical coding-theoretic bounds available for the $\lambda$-subfield metric by deriving Griesmer-, Elias- and Johnson-type bounds. 
Together with the previously known Singleton-, Plotkin-, sphere-packing-, and Gilbert-Varshamov-type bounds, this provides a broad analogue of the classical theory for the Hamming metric and the rank metric.

Our results show that many fundamental concepts from classical coding theory admit natural analogues in the $\lambda$-subfield metric. At the same time, several questions remain open. 
In particular, explicit constructions of M$\lambda$D codes beyond the currently known families,  and efficient decoding algorithms remain largely unexplored. We hope that the results presented here provide a foundation for further investigations into the $\lambda$-subfield metric and its applications in coding theory and quantum error correction.
\section*{Acknowledgments}
 Violetta Weger’s work was supported by the Technical University of Munich—Institute for Advanced Study.
 
Violetta Weger is funded by the European Union ERC-StG CODES - 101304377. Views and opinions expressed are however those of the author(s) only and do not necessarily reflect those of the European Union or the European Research Council. Neither the European Union nor the granting authority can be held responsible for them.

\bibliographystyle{plain}
\bibliography{biblio}

\end{document}